\input harvmac
\input amssym 
\input epsf 
 
\def\CJ{{\cal J}}

\def\HA{{\hat A}} 
\def\HF{{\hat F}} 
\def\HJ{{\hat J}} 
 
\def\Th{{\tilde h}} 
\def\Tk{{\tilde k}} 
 
\def\Tz{{\tilde z}} 
\def\TZ{{\tilde Z}} 
\def\Tkappa{{\tilde \kappa}} 
\def\vecX{{\overrightarrow X}} 
\def\vecx{{\overrightarrow x}} 
\def\vecy{{\overrightarrow y}} 
\def\veca{{\overrightarrow a}}

\lref\AK{ 
  O.~Aharony and D.~Kutasov, 
  ``Holographic Duals of Long Open Strings,'' 
  arXiv:0803.3547 [hep-th]. 
  %%CITATION = ARXIV:0803.3547;%% 
} 
\lref\AN{ 
  G.~S.~Adkins and C.~R.~Nappi, 
  ``The Skyrme Model with Pion Masses,'' 
  Nucl.\ Phys.\  B {\bf 233} (1984) 109. 
  %%CITATION = NUPHA,B233,109;%% 
} 
\lref\ASY{ 
  O.~Aharony, J.~Sonnenschein and S.~Yankielowicz, 
  ``A Holographic Model of Deconfinement and Chiral Symmetry Restoration,'' 
  Annals Phys.\  {\bf 322} (2007) 1420 
  [arXiv:hep-th/0604161]. 
  %%CITATION = APNYA,322,1420;%% 
} 
\lref\BISY{ 
  A.~Brandhuber, N.~Itzhaki, J.~Sonnenschein and S.~Yankielowicz, 
  ``Baryons from Supergravity,'' 
  JHEP {\bf 9807} (1998) 020 
  [arXiv:hep-th/9806158]. 
  %%CITATION = JHEPA,9807,020;%% 
} 
\lref\BLL{ 
  O.~Bergman, G.~Lifschytz and M.~Lippert, 
  ``Holographic Nuclear Physics,'' 
  JHEP {\bf 0711} (2007) 056 
  [arXiv:0708.0326 [hep-th]]. 
  %%CITATION = JHEPA,0711,056;%% 
} 
\lref\BLLi{ 
  O.~Bergman, G.~Lifschytz and M.~Lippert, 
  ``Magnetic Properties of Dense Holographic QCD,'' 
  arXiv:0806.0366 [hep-th]. 
  %%CITATION = ARXIV:0806.0366;%% 
} 
\lref\BPST{ 
  A.~A.~Belavin, A.~M.~Polyakov, A.~S.~Shvarts and Yu.~S.~Tyupkin, 
  ``Pseudoparticle Solutions of the Yang-Mills Equations,'' 
  Phys.\ Lett.\  B {\bf 59} (1975) 85. 
  %%CITATION = PHLTA,B59,85;%% 
} 
\lref\BSS{ 
  O.~Bergman, S.~Seki and J.~Sonnenschein, 
  ``Quark Mass and Condensate in HQCD,'' 
  JHEP {\bf 0712} (2007) 037 
  [arXiv:0708.2839 [hep-th]]. 
  %%CITATION = JHEPA,0712,037;%% 
} 
\lref\CGS{ 
  C.~G.~Callan, A.~Guijosa and K.~G.~Savvidy, 
  ``Baryons and String Creation from the Fivebrane Worldvolume Action,'' 
  Nucl.\ Phys.\  B {\bf 547} (1999) 127 
  [arXiv:hep-th/9810092]. 
  %%CITATION = NUPHA,B547,127;%% 
} 
\lref\CGST{ 
  C.~G.~Callan, A.~Guijosa, K.~G.~Savvidy and O.~Tafjord, 
  ``Baryons and Flux Tubes in Confining Gauge Theories from Brane Actions,'' 
  Nucl.\ Phys.\  B {\bf 555} (1999) 183 
  [arXiv:hep-th/9902197]. 
  %%CITATION = NUPHA,B555,183;%% 
} 
\lref\CKP{ 
  R.~Casero, E.~Kiritsis and A.~Paredes, 
  ``Chiral Symmetry Breaking as Open String Tachyon Condensation,'' 
  Nucl.\ Phys.\  B {\bf 787} (2007) 98 
  [arXiv:hep-th/0702155]. 
  %%CITATION = NUPHA,B787,98;%% 
} 
\lref\CM{ 
  C.~G.~Callan and J.~M.~Maldacena, 
  ``Brane Dynamics from the Born-Infeld Action,'' 
  Nucl.\ Phys.\  B {\bf 513} (1998) 198 
  [arXiv:hep-th/9708147]. 
  %%CITATION = NUPHA,B513,198;%% 
} 
\lref\CPS{ 
  R.~Casero, A.~Paredes and J.~Sonnenschein, 
  ``Fundamental Matter, Meson Spectroscopy and Non-critical String / Gauge 
  Duality,'' 
  JHEP {\bf 0601} (2006) 127
  [arXiv:hep-th/0510110]. 
  %%CITATION = JHEPA,0601,127;%% 
} 
\lref\DN{ 
  A.~Dhar and P.~Nag, 
  ``Sakai-Sugimoto Model, Tachyon Condensation and Chiral Symmetry Breaking,'' 
  JHEP {\bf 0801} (2008) 055 
  [arXiv:0708.3233 [hep-th]]. 
  %%CITATION = JHEPA,0801,055;%% 
} 
\lref\GO{ 
  D.~J.~Gross and H.~Ooguri, 
  ``Aspects of Large N Gauge Theory Dynamics as Seen by String Theory,'' 
  Phys.\ Rev.\  D {\bf 58} (1998) 106002 
  [arXiv:hep-th/9805129]. 
  %%CITATION = PHRVA,D58,106002;%% 
} 
\lref\Ha{ 
  K.~Hashimoto, 
  ``Holographic Nuclei,'' 
  arXiv:0809.3141 [hep-th]. 
  %%CITATION = ARXIV:0809.3141;%% 
} 
\lref\HHLY{ 
  K.~Hashimoto, T.~Hirayama, F.~L.~Lin and H.~U.~Yee, 
  ``Quark Mass Deformation of Holographic Massless QCD,'' 
  JHEP {\bf 0807} (2008) 089 
  [arXiv:0803.4192 [hep-th]]. 
  %%CITATION = JHEPA,0807,089;%% 
} 
\lref\HLPY{ 
  D.~K.~Hong, K.~M.~Lee, C.~Park and H.~U.~Yee, 
  ``Holographic Monopole Catalysis of Baryon Decay,'' 
  JHEP {\bf 0808} (2008) 018 
  [arXiv:0804.1326 [hep-th]]. 
  %%CITATION = JHEPA,0808,018;%% 
} 
\lref\HM{ 
  H.~Hata and M.~Murata, 
  ``Baryons and the Chern-Simons Term in Holographic QCD with Three Flavors,'' 
  Prog.\ Theor.\ Phys.\  {\bf 119} (2008) 461 
  [arXiv:0710.2579 [hep-th]]. 
  %%CITATION = PTPKA,119,461;%% 
} 
\lref\HMY{ 
  H.~Hata, M.~Murata and S.~Yamato, 
  ``Chiral Currents and Static Properties of Nucleons in Holographic QCD,'' 
  arXiv:0803.0180 [hep-th]. 
  %%CITATION = ARXIV:0803.0180;%% 
} 
\lref\HRYY{ 
  D.~K.~Hong, M.~Rho, H.~U.~Yee and P.~Yi, 
  ``Chiral Dynamics of Baryons from String Theory,'' 
  Phys.\ Rev.\  D {\bf 76} (2007) 061901 
  [arXiv:hep-th/0701276]. 
  %%CITATION = PHRVA,D76,061901;%% 
} 
\lref\HRYYi{ 
  D.~K.~Hong, M.~Rho, H.~U.~Yee and P.~Yi, 
  ``Dynamics of Baryons from String Theory and Vector Dominance,'' 
  JHEP {\bf 0709} (2007) 063  
  [arXiv:0705.2632 [hep-th]]. 
  %%CITATION = JHEPA,0709,063;%% 
} 
\lref\HSS{ 
  K.~Hashimoto, T.~Sakai and S.~Sugimoto, 
  ``Holographic Baryons : Static Properties and Form Factors from Gauge/String 
  Duality,'' 
  arXiv:0806.3122 [hep-th]. 
  %%CITATION = ARXIV:0806.3122;%% 
} 
\lref\HSSY{ 
  H.~Hata, T.~Sakai, S.~Sugimoto and S.~Yamato, 
  ``Baryons from Instantons in Holographic QCD,'' 
  arXiv:hep-th/0701280. 
  %%CITATION = HEP-TH/0701280;%% 
} 
\lref\Imaa{ 
  A.~Imaanpur, 
  ``On Instantons in Holographic QCD,'' 
  arXiv:0705.0496 [hep-th]. 
  %%CITATION = ARXIV:0705.0496;%% 
} 
\lref\Imam{ 
  Y.~Imamura, 
  ``String Junctions and Their Duals in Heterotic String Theory,'' 
  Prog.\ Theor.\ Phys.\  {\bf 101} (1999) 1155
  [arXiv:hep-th/9901001]. 
  %%CITATION = PTPKA,101,1155;%% 
} 
\lref\KSi{ 
  S.~Kuperstein and J.~Sonnenschein, 
  ``Non-critical Supergravity ($d > 1$) and Holography,'' 
  JHEP {\bf 0407} (2004) 049 
  [arXiv:hep-th/0403254]. 
  %%CITATION = JHEPA,0407,049;%% 
} 
\lref\KSii{ 
  S.~Kuperstein and J.~Sonnenschein, 
  ``Non-critical, Near Extremal $AdS_6$ Background as a Holographic Laboratory 
  of Four Dimensional YM Theory,'' 
  JHEP {\bf 0411} (2004) 026 
  [arXiv:hep-th/0411009]. 
  %%CITATION = JHEPA,0411,026;%% 
} 
\lref\KSZ{ 
  K.~Y.~Kim, S.~J.~Sin and I.~Zahed, 
  ``The Chiral Model of Sakai-Sugimoto at Finite Baryon Density,'' 
  JHEP {\bf 0801} (2008) 002 
  [arXiv:0708.1469 [hep-th]]. 
  %%CITATION = JHEPA,0801,002;%% 
} 
\lref\KSZi{ 
  K.~Y.~Kim, S.~J.~Sin and I.~Zahed, 
  ``Dense Holographic QCD in the Wigner-Seitz Approximation,'' 
  JHEP {\bf 0809} (2008) 001 
  [arXiv:0712.1582 [hep-th]]. 
  %%CITATION = JHEPA,0809,001;%% 
} 
\lref\KSZii{ 
  K.~Y.~Kim, S.~J.~Sin and I.~Zahed, 
  ``Dense and Hot Holographic QCD: Finite Baryonic E Field,'' 
  JHEP {\bf 0807} (2008) 096 
  [arXiv:0803.0318 [hep-th]]. 
  %%CITATION = JHEPA,0807,096;%% 
} 
\lref\KZ{ 
  K.~Y.~Kim and I.~Zahed, 
  ``Electromagnetic Baryon Form Factors from Holographic QCD,'' 
  JHEP {\bf 0809} (2008) 007 
  [arXiv:0807.0033 [hep-th]]. 
  %%CITATION = JHEPA,0809,007;%% 
} 
\lref\KZSV{ 
  M.~Kruczenski, L.~A.~P.~Zayas, J.~Sonnenschein and D.~Vaman, 
  ``Regge Trajectories for Mesons in the Holographic Dual of Large-$N_c$ QCD,'' 
  JHEP {\bf 0506} (2005) 046 
  [arXiv:hep-th/0410035]. 
  %%CITATION = JHEPA,0506,046;%% 
} 
\lref\Ma{ 
  J.~M.~Maldacena, 
  ``The Large N Limit of Superconformal Field Theories and Supergravity,'' 
  Adv.\ Theor.\ Math.\ Phys.\  {\bf 2} (1998) 231 
  [Int.\ J.\ Theor.\ Phys.\  {\bf 38} (1999) 1113] 
  [arXiv:hep-th/9711200]. 
  %%CITATION = IJTPB,38,1113;%% 
} 
\lref\MB{ 
  O.~Mintkevich and N.~Barnea, 
  ``Wave Function for No-core Effective Interaction Approaches,'' 
  Phys.\ Rev.\  C {\bf 69} (2004) 044005. 
  %%CITATION = PHRVA,C69,044005;%% 
} 
\lref\MS{ 
  V.~Mazo and J.~Sonnenschein, 
  ``Non Critical Holographic Models of the Thermal Phases of QCD,'' 
  JHEP {\bf 0806} (2008) 091 
  [arXiv:0711.4273 [hep-th]]. 
  %%CITATION = JHEPA,0806,091;%% 
}
\lref\NSK{
  K.~Nawa, H.~Suganuma and T.~Kojo,
  ``Baryons in Holographic QCD,''
  Phys.\ Rev.\  D {\bf 75} (2007) 086003 
  [arXiv:hep-th/0612187]; 
  %%CITATION = PHRVA,D75,086003;%%
}
\lref\NSKi{
  K.~Nawa, H.~Suganuma and T.~Kojo,
  ``Brane-induced Skyrmion on $S^3$: baryonic matter in holographic QCD,''
  arXiv:0810.1005 [hep-th].
  %%CITATION = ARXIV:0810.1005;%%
}
\lref\PDG{ 
  W.~M.~Yao {\it et al.}  [Particle Data Group], 
  ``Review of Particle Physics,'' 
  J.\ Phys.\ G {\bf 33} (2006) 1. 
  %%CITATION = JPHGB,G33,1;%% 
} 
\lref\PW{
  A.~Pomarol and A.~Wulzer,
  ``Baryon Physics in Holographic QCD,''
  arXiv:0807.0316 [hep-ph].
  %%CITATION = ARXIV:0807.0316;%%
}
\lref\PY{ 
  J.~Park and P.~Yi, 
  ``A Holographic QCD and Excited Baryons from String Theory,'' 
  JHEP {\bf 0806} (2008) 011 
  [arXiv:0804.2926 [hep-th]]. 
  %%CITATION = JHEPA,0806,011;%% 
} 
\lref\PZ{ 
  K.~Peeters and M.~Zamaklar, private communication. 
} 
\lref\RSVW{ 
  M.~Rozali, H.~H.~Shieh, M.~Van Raamsdonk and J.~Wu, 
  ``Cold Nuclear Matter in Holographic QCD,'' 
  JHEP {\bf 0801} (2008) 053 
  [arXiv:0708.1322 [hep-th]]. 
  %%CITATION = JHEPA,0801,053;%% 
} 
\lref\SS{ 
  T.~Sakai and S.~Sugimoto, 
  ``Low Energy Hadron Physics in Holographic QCD,'' 
  Prog.\ Theor.\ Phys.\  {\bf 113} (2005) 843 
  [arXiv:hep-th/0412141]. 
  %%CITATION = PTPKA,113,843;%% 
} 
\lref\SSi{ 
  T.~Sakai and S.~Sugimoto, 
  ``More on a Holographic Dual of QCD,'' 
  Prog.\ Theor.\ Phys.\  {\bf 114} (2006) 1083 
  [arXiv:hep-th/0507073]. 
  %%CITATION = PTPKA,114,1083;%% 
} 
\lref\SeS{ 
  Y.~Seo and S.~J.~Sin, 
  ``Baryon Mass in Medium with Holographic QCD,'' 
  JHEP {\bf 0804} (2008) 010 
  [arXiv:0802.0568 [hep-th]]. 
  %%CITATION = JHEPA,0804,010;%% 
} 
\lref\Wi{ 
  E.~Witten, 
  ``Baryons and Branes in Anti de Sitter Space,'' 
  JHEP {\bf 9807} (1998) 006 
  [arXiv:hep-th/9805112]. 
  %%CITATION = JHEPA,9807,006;%% 
} 
\lref\Wii{ 
  E.~Witten, 
  ``Anti-de Sitter Space, Thermal Phase Transition, and Confinement in Gauge 
  Theories,'' 
  Adv.\ Theor.\ Math.\ Phys.\  {\bf 2} (1998) 505 
  [arXiv:hep-th/9803131]. 
  %%CITATION = 00203,2,505;%% 
}

\Title{\vbox{\baselineskip12pt 
    \hbox{TAUP-2885/08}\hbox{\tt arXiv:0810.1633}}} 
{Comments on Baryons in Holographic QCD} 
 
\centerline{Shigenori Seki\footnote{$^*$}{\tt sekish@post.tau.ac.il} and 
    Jacob Sonnenschein\footnote{$^\dagger$}{\tt cobi@post.tau.ac.il}} 
\bigskip 
\centerline{\it The Reymond and Beverly Sackler School of Physics and Astronomy} 
\centerline{\it Faculty of Exact Sciences, Tel Aviv University, Ramat Aviv 69978, Israel} 
 
%if too many authors for abstract on same page, say   \vfill\eject\pageno0 
 
\vskip .3in 
 
\centerline{\bf Abstract} 

We generalize the description of baryons 
as instantons of Sakai-Sugimoto model to the case where the 
flavor branes are non-anti-podal. 
The later corresponds to quarks with a ``string endpoint  mass''. 
We show that the baryon vertex is located 
on the flavor branes and hence the generalized baryons also 
associate with  instantons. 
We calculate the baryon mass spectra, 
the isoscalar and axial mean square radii, 
the isoscalar and isovector magnetic moments 
and the axial coupling as a function of 
the mass scale  $M_{\rm KK}$ and the location $\zeta$ of the tip of 
U-shaped flavor D8-branes. 
We determine the values of $M_{\rm KK}$ and $\zeta$ 
from a best fit comparison with the experimental data. 
The later comes out to be in a forbidden region, 
which may indicate that the incorporation of baryons 
in Sakai-Sugimoto model has to be modified. 
We discuss the analogous baryons in a non-critical gravity model. 
A brief comment on the single flavor case $(N_f=1)$ is also made.

\Date{9 October 2008} 

%%%%%%%%%%%%%%%%%%%%%%%%%%%%%% 
\newsec{Introduction}
 
Baryons were incorporated into the $AdS_5\times S^5$ model 
in \refs{\Wi,\GO} via a D5-brane wrapping the $S^5$ with 
$N_c$ strings attached to it and ending up at the boundary. 
The strings are needed to cancel an $N_c$ charge 
in the world-volume of the wrapped brane 
that follows from the RR flux of the background. 
This object which is the dual of an external baryon, 
namely with infinitely heavy quarks was further discussed 
in \refs{\Imam\CGS{--}\CGST} 
and was generalized also to confining backgrounds \BISY\ 
where it was found that their energy was linear 
in $N_c$ and in the ``size'' of the baryon on the boundary. 
 
A realization of a dynamical baryon has become possible once flavor 
probe branes were added to holographic models. 
A prototype of such a model is Sakai-Sugimoto (SS) model \SS. 
This model is based on placing a stack of $N_f$ probe D8-branes 
and a stack of $N_f$ probe anti-D8-branes 
connected in a U-shaped cigar profile, 
into the model of \Wii\ of near extremal D4-branes.  
The baryon vertex is immersed in the probe brane at the tip of the cigar. 
In \HSSY\ it was shown that the baryon corresponds to an instanton of 
the five-dimensional effective $U(N_f=2)$ gauge theory. 
The physical properties of this baryon were analyzed in several papers 
\refs{\HRYY\Imaa\HRYYi\BLL\RSVW\KSZ\HM\KSZi\SeS\HMY\KSZii\HLPY\PY\BLLi\KZ{--}\Ha}\foot{
The different approach for the baryons in SS model 
has been studied by \refs{\NSK,\NSKi}.}. 
These include in particular the mass, size, mass splitting, 
the mean square radii, magnetic moments, various couplings and more. 
A comparison with experimental data reveals an agreement similar, 
or even better, than the one found in the Skyrme model \AN. 
In spite of this success the baryons of the model of \HSSY\ suffer 
from several problems. The size of the baryon is proportional to 
$\lambda^{-1/2}$ where $\lambda$ is the four-dimensional 't Hooft 
parameter.  
Since the gravitational holographic model is valid only 
in the large $\lambda$ limit, 
this implies that stringy corrections have to be taken into account. 
Another drawback of the model is that the scale of the system 
associated with the baryonic structure is 
roughly half the one needed to fit to the mesonic data\foot{
Ref.~\PW\ has shown that this problem is substantially improved in the
AdS/QCD model. }.

SS model has a generalization \ASY, where the location 
of the probe branes in the compactified direction is not 
anti-podal, or differently stating the tip of the probe brane 
is at a radial location $u_0 > u_{\rm KK}$ where $u_{\rm KK}$ 
is the minimal value of the radial direction of the background. 
The difference between the two cases is drawn in fig.~1. 
\bigskip 
\vbox{\centerline{\epsfbox{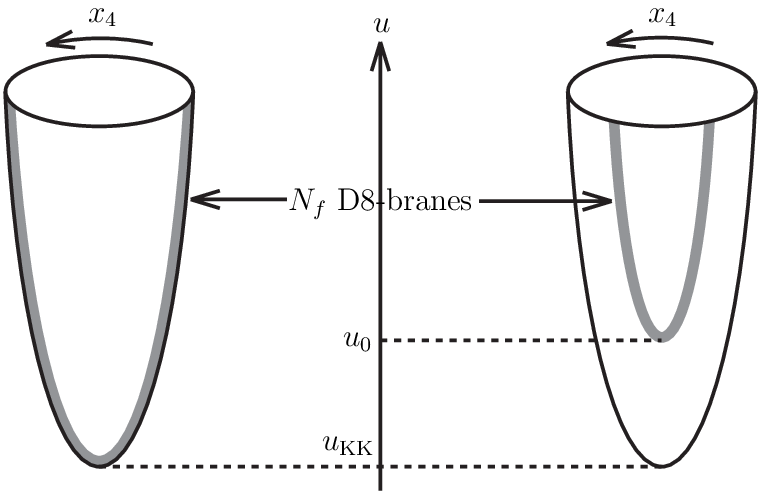}} 
\centerline{\fig\figcigar{} The probe D8-branes in the cigar background}} 
\bigskip\nobreak\noindent 
The non-anti-podal case is in fact a family of models characterized 
by the separation distance $L$ or a ``string endpoint mass'' 
of the quark \KZSV\foot{For attempts to 
introduce the QCD or current algebra mass, see 
\refs{\CKP\BSS\DN\AK{--}\HHLY}.}. 
A natural question to ask is how do the properties of the baryon 
depend on the additional parameter and in particular 
whether the problems mentioned above in the context of 
the anti-podal case can be circumvented.  
This is the main goal of this paper. 
As a first step we address the question of where the baryon vertex is located in the generalized setup. 
We show that in the confining phase it is again immersed in the probe brane. 
In the deconfining phase above a certain critical temperature, 
the baryon vertex falls into the ``black hole'' 
and thus the baryon is dissolved. 
The main part of this paper includes a repetition 
of the calculations performed in \refs{\HSSY,\HSS} 
of the properties of the baryons now made in the generalized setup 
with non-trivial stringy mass namely a non-anti-podal configuration. 
The expressions for the mass spectra, mean radii, magnetic moments 
and couplings are derived as a function of the scale and the parameter 
which measures the deviation from the anti-podal configuration. 
It has turned out that the generalized setup does not resolve 
the problem of the size of the baryon. 
We have found that the data can be fit with the same scale 
that governs the mesonic spectra provided the location 
of the probe brane is in an unphysical location 
``below the tip of the cigar''. 
It seems to us that this is an indication of a problem 
of the baryonic setup of SS model.

We also analyze the baryons of the non-critical model \KSii\ based on 
the incorporation of $N_f$ probe D4-branes into the background of a 
near extremal D4-branes residing in six dimensions. 
It is shown that the problem of the small size of the baryon 
is avoided in this model. 
We also setup the stage for the open problem of the baryons 
of a single flavor brane namely $N_f=1$.

The paper is organized as follows. 
After this introduction we describe the general setup 
of the non-anti-podal SS model. 
In Section 3 we analyze the baryonic configuration in the 
generalized setup 
and determine that the location of the baryon vertex 
is on the flavor brane. 
Section 4 is devoted to a detailed analysis of the baryon properties 
following \refs{\HSSY,\HSS} in the non-anti-podal geometry. 
The values of the scale and the location of the flavor brane 
that fit the data in an optimal way are determined. 
We then present the open question of the baryon 
for a single flavor case. 
Section 6 presents an analysis similar to the one in Section 4 
but in the context of a non-critical six-dimensional model. 
We end with a short summary, list of conclusions and open questions. 
Appendix includes the computations of the location 
of the baryon vertex in the general case of D$p$-brane background 
with D$(8-p)$-branes wrapping an $S^{8-p}$ cycle.

%%%%%%%%%%%%%%%%%%%%%%%%%%%%%% 
\newsec{The general setup of the non-anti-podal Sakai-Sugimoto model} 
 
SS model \SS\ is a system which consists of 
$N_c$ coincident color D4-branes and $N_f$ coincident flavor D8-branes. 
When $N_c$ is large, the D4-branes are regarded as the background, 
of which metric is given by 
\eqn\Dfourbg{\eqalign{ 
&ds^2 = \biggl({u 
\over R}\biggr)^{3 \over 2} 
    \bigl[\eta_{\mu\nu}dx^\mu dx^\nu + f(u) dx_4^2\bigr] 
    +\biggl({R \over u}\biggr)^{3 \over 2}\biggl[{du^2 \over f(u)} + u^2d\Omega_4^2\biggr] , \cr 
&e^\phi = g_s\biggl({u \over R}\biggr)^{3 \over 4} ,\quad 
F_{(4)} = {2\pi N_c \over V_4}\epsilon_4 ,\quad 
R^3 := \pi g_s N_c \l_s^3 ,\quad 
f(u) := 1- \biggl({u_{\rm KK} \over u}\biggr)^3 , 
}} 
where $\eta_{\mu\nu} = {\rm diag}(-1,1,1,1)$. 
The volume of unit four sphere $V_4$ is equal to $8\pi^2/3$. 
The $x_4$ direction is compactified by the circle with the period 
\eqn\xfourper{ 
\delta x_4 = {4\pi R^{3 \over 2} \over 3u_{\rm KK}^{1 \over 2}} . 
} 
This period is determined so that the singularity at the tip $u=u_{\rm KK}$ 
is excluded. 
Then the Kaluza-Klein mass scale $M_{\rm KK}$ becomes 
\eqn\KKmass{ 
M_{\rm KK} := {2\pi \over \delta x_4} 
    = {3 u_{\rm KK}^{1 \over 2} \over 2 R^{3 \over 2}} . 
} 
 
The flavor D8-branes are realized as the probe 
in the D4-branes' background \Dfourbg. 
The action of the coincident D8-branes consists of the two parts, 
\eqn\Deightact{ 
S_{\rm D8} = S_{\rm DBI} + S_{\rm CS} . 
} 
$S_{\rm DBI}$ is the Dirac-Born-Infeld (DBI) action, 
\eqn\DeightDBI{ 
S_{\rm DBI} = T_8\int d^9x\, e^{-\phi} \sqrt{-\det(g_{MN} + 2\pi\alpha' \CF_{MN})} , 
} 
where the D8-brane's tension is denoted by $T_8 = (2\pi)^{-8}l_s^{-9}$. 
The induced metric $g_{MN}$ is computed from \Dfourbg, 
\eqn\DeightIndmet{ 
ds_{\rm D8}^2 = \biggl({u \over R}\biggr)^{3 \over 2}\eta_{\mu\nu}dx^\mu dx^\nu 
    + \Biggl[\biggl({u \over R}\biggr)^{3 \over 2}f(u) 
    + \biggl({R \over u}\biggr)^{3 \over 2}{u'^2 \over f(u)}\Biggr]dx_4^2 
    +\biggl({R \over u}\biggr)^{3 \over 2}u^2d\Omega_4^2 , 
} 
where $u'$ denotes $du/dx_4$. $\CF$ is a $U(N_f)$ gauge field strength 
on the worldvolume of the D8-branes. 
The $U(N_f)$ gauge field $\CA$ has also Chern-Simon action $S_{\rm CS}$, 
\eqn\NfCSact{ 
S_{\rm CS} = {N_c \over 24\pi^2} \int \tr\biggl(\CA\CF^2-{i \over 2}\CA^3\CF-{1 \over 10}\CA^5\biggr) . 
} where the integral is now a five-dimensional one.

We shall study the shape of the D8-branes by the analyses of 
the classical solution of \Deightact\ without the gauge fields. 
In terms of \DeightIndmet, the DBI action \DeightDBI\ is written down 
\eqn\DBIo{ 
S_{\rm DBI} = {T_8 \Omega_4 \over g_s} \int d^4x dx_4\, 
    u^4\sqrt{f(u) + \biggl({R \over u}\biggr)^3 {u'^2 \over f(u)}} =: S_0[u(x_4)] . 
} 
Since the Hamiltonian calculated from this action is the function 
of only $u$, we can put the Hamiltonian constraint, 
\eqn\Hamiconst{ 
{u^4 f(u) \over \sqrt{f(u) + \bigl({R \over u}\bigr)^3 {u'^2 \over f(u)}}} 
= {\rm constant} = u_0^4 \sqrt{f(u_0)} , 
} 
where we used $u(0) = u_0$ and $u'(0) = 0$. 
Note that $u_0 \geq u_{\rm KK}$. The Hamiltonian constraint is 
rewritten as 
\eqn\uxfour{ 
{du \over dx_4} = \pm \biggl({u \over R}\biggr)^{3 \over 2}f(u) 
    \sqrt{{u^8f(u) \over u_0^8 f(u_0)}-1} . 
} 
The solution of this equation implies that the D8-branes are U-shape 
in the cigar geometry expanded by the $(u,x_4)$ coordinates (see also fig.~1). 
The boundary value $x_4(u=\infty) := L/2$ is evaluated from \uxfour 
\eqn\Luzero{ 
L = \int_{-L/2}^{L/2} dx_4 = 2\int_{u_0}^\infty {du \over |u'|} 
= 2\int_{u_0}^\infty \biggl({R \over u}\biggr)^{3 \over 2}{1 \over f(u)\sqrt{{u^8f(u) \over u_0^8 f(u_0)}-1}} du. 
} 
$L$ denotes the separation along the $x_4$ direction between the D8-branes 
at $u=\infty$. 
The equation \Luzero\ relates the parameter $u_0$ at the IR ($u=u_0$) with 
$L$ at the UV ($u=\infty$). When $u_0$ is equal to $u_{\rm KK}$, 
in other words, $L=\delta x_4/2$, the D8-branes are located at the 
anti-podal positions on the circular $x_4$ direction. 
This anti-podal case is the original SS model \refs{\SS,\SSi}.

%%%%%%%%%%%%%%%%%%%%%%%%%%%%%% 
\newsec{The baryon configuration in the genralized Sakai-Sugimoto
model}
 
The external baryon of the model of \Wi\ was explored in 
\BISY. It is composed from a baryon vertex 
 which is  a D4-brane wrapped on $S^4$ and $N_c$ fundamental strings 
stretched between this D4-brane and the boundary. 
A dynamical baryon in the model of \SS\ differs from 
the external one in that the strings end on the probe flavor D8-branes 
and not on the boundary. 
%This is analogous to the baryon vertex suggested by 
%\refs{\CM,\CGS,\CGST}. 
The leading order action, which is the sum of the action of 
the D4-brane and the action of the $N_c$ strings, takes the form 
$$ 
S = -T_4 \int dt d\Omega_4 e^{-\phi}\sqrt{-\det g_{\rm D4}} 
    -N_c T_f \int dt du \sqrt{-\det g_{\rm string}} =: -\int dt\, E . 
$$ 
where $E$ is the energy density and 
$$ 
T_4 = (2\pi)^{-4} l_s^{-5} ,\quad T_f = (2\pi)^{-1} l_s^{-2} . 
$$

In a similar  way one can consider the baryonic D$(8-p)$-brane 
wrapped on the $(8-p)$-dimensional sphere in the $N_c$ D$p$-branes' 
background. This baryonic D-brane is regarded as the baryon vertex 
in the $p$-dimensional QCD-like theory. This analysis is presented 
in Appendix A. 

The idea now is to find the location of the baryon vertex 
from the requirement of minimizing the energy. 
The energy as a function of the location of the baryon vertex 
will be calculated for the two distinct systems 
of the confining background and the deconfining one.

\subsec{Confinement phase} 

The confining background is given by \Dfourbg. 
Substituting this into the expression of the energy, 
we find 
$$\eqalign{ 
E(u_B;u_0) &= {N_c \over 2\pi l_s^2}\biggl[ 
    {1 \over 3}u_B + \int^{u_0}_{u_B}du {1 \over \sqrt{f(u)}}\biggr] 
    =: {N_c u_{\rm KK} \over 2\pi l_s^2} {\cal E}_{\rm conf}(x;x_0) , \cr 
&{\cal E}_{\rm conf}(x;x_0) = {1 \over 3}x 
    +\int^{x_0}_x {dy \over \sqrt{1-y^{-3}}} , 
}$$ 
where $x:=u_B/u_{\rm KK}$ and $x_0 := u_0/u_{\rm KK}$ , 
the valid range of $x$ is $1 \leq x \leq x_0$ (see fig.~2). 
\bigskip 
\vbox{\centerline{\epsfbox{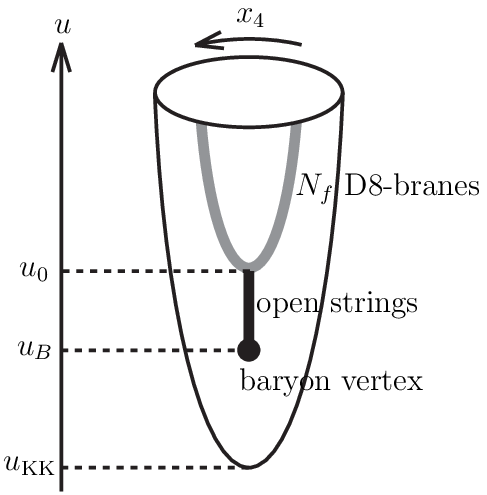}} 
\centerline{\fig\figconfBV{} The baryon vertex in the confinement phase}} 
\bigskip\nobreak\noindent 
Since ${\cal E}_{\rm conf}(x;x_0)$ is a monotonically 
decreasing function of $x$, 
the energy $E$ becomes minimum at $x = x_0$. 
 
The meaning of this result is that like the anti-podal case also 
for the generalized case where $x_0 (= u_0/u_{\rm KK}) \neq 1$ 
the baryon vertex is immersed inside the flavor probe branes.
As was mentioned this is only a leading order calculation. 
It can be improved by adding the energy associated 
with the deformation of the wrapped brane due to the strings \CM, 
and by relaxing the assumption that the strings stretch only 
along the radial direction. 
We believe that these improvements would not change the conclusion 
that the baryon vertex is located on the probe branes.

\subsec{Deconfinement phase} 

Next we study the location of the baryon vertex 
in the deconfining phase. 
The difference in the background metric is 
that now the thermal factor is dressing 
the compactified Euclidean time direction, 
and we replace the scale with the one related to 
the temperature $u_T$. 
Since the background metric in this phase reads 
$$\eqalign{ 
&ds^2 = \biggl({u \over R}\biggr)^{3 \over 2} 
    \bigl[f_T(u)dt^2 +\delta_{ij}dx^i dx^j + dx_4^2\bigr] 
    +\biggl({R \over u}\biggr)^{3 \over 2}\biggl[{du^2 \over f_T(u)} + u^2d\Omega_4^2\biggr] , \cr 
&f_T(u) := 1- \biggl({u_T \over u}\biggr)^3 , 
}$$ 
the corresponding energy can be evaluated 
$$\eqalign{ 
E(u_B;u_0) &= {N_c \over 2\pi l_s^2}\biggl[{1 \over 
3}u_B\sqrt{f_T(u_B)} +(u_0 - u_B)\biggr] 
        =: {N_c u_T \over 2\pi l_s^2} {\cal E}_{\rm deconf}(x;x_0) , \cr 
&{\cal E}_{\rm deconf}(x;x_0) = {1 \over 3}x\sqrt{1 - {1 \over x^3}} 
+ (x_0 - x), }$$ 
where $x:= u_B/u_T$, $x_0 := u_0/u_T$ and $1 \leq x \leq x_0$. 
\bigskip 
\vbox{\centerline{\epsfbox{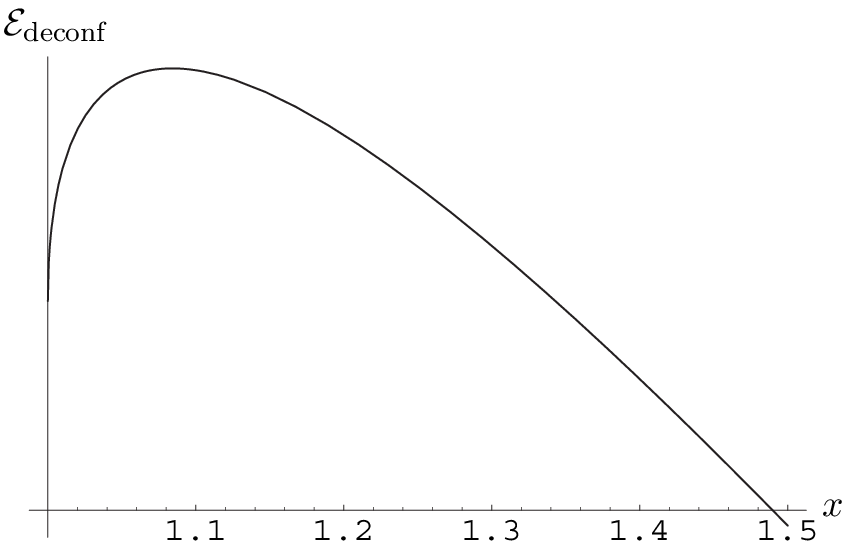}} 
\centerline{\fig\figdeconf{${\cal E}_{\rm deconf}(x)$} ${\cal 
E}_{\rm deconf}(x)$}} 
\bigskip\nobreak\noindent 
The energy (\figdeconf) has a maximum at 
$$ 
x = \biggl({5+3\sqrt{3} \over 8}\biggr)^{1 \over 3} =: x_{\rm max} . 
$$ 
$x_{\rm max}$ is approximately equal to 1.08422. 
We are also interested in the critical value $x_{\rm cr}$ 
which satisfies 
$$ 
E(1;x_0) = E(x_{\rm cr};x_0) . 
$$ 
$x_{\rm cr}$ can be analitically calculated, 
\eqn\deconfcr{ 
x_{\rm cr} = {5 +\sqrt{33} \over 8} \approx 1.34307 . 
} 
If $x_0 > x_{\rm cr}$, then the energy becomes minimum 
at $x = x_0$ and the baryon vertex can exist 
at the tip of the U-shaped flavor D8-brane (fig.~4(a)). 
On the other hand, if $x_0 < x_{\rm cr}$, 
then the energy becomes minimum at $x = 1$, that is to say, 
the baryon vertex falls down into the black hole (fig.~4(b)). 
\bigskip 
\vbox{\centerline{\epsfbox{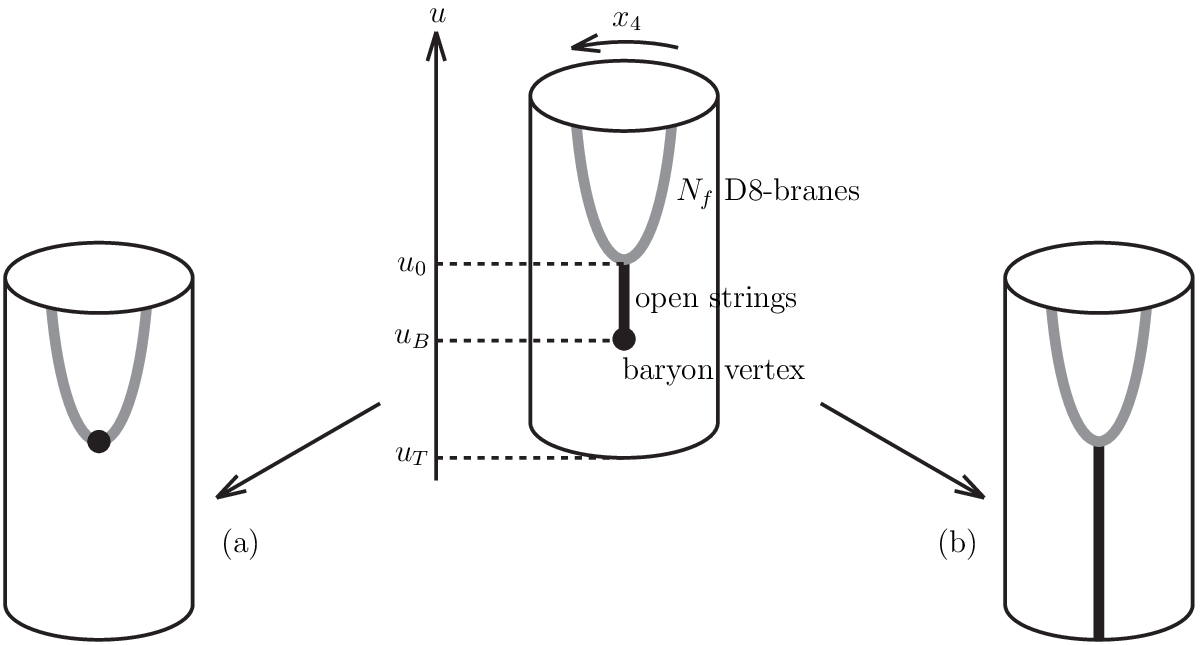}} 
\centerline{\fig\figdeconfBV{} The baryon vertex in the deconfinement phase}} 
\bigskip\nobreak\noindent 
The physical meaning of the picture is that for 
temperatures lower than a critical temperature, 
which is higher than the temperature of 
the confinement/deconfinement phase transition, 
the baryon vertex will be in the flavor brane just as 
in the zero temperature case. 
However, for higher temperature the baryon is dissolved 
via falling into the black hole and 
becoming $N_c$ deconfined quarks.

%%%%%%%%%%%%%%%%%%%%%%%%%%%%%% 
\newsec{Baryons as instantons in non-anti-podal Sakai-Sugimoto model} 

Once we found that the baryon vertex is immersed 
inside the probe flavor branes, 
to extract the properties of the baryons 
we have to repeat the computations done in \refs{\HSSY, \HSS} 
in the setup descibed in Section 2 
rather than in the anti-podal geometry. 
 
We turn on the $U(N_f)$ gauge fields 
as the perturbation around the classical solution $\CA = 0$ 
discussed in Section 2. 
The DBI action \DeightDBI\ is expanded with respect to the gauge field, 
$$ 
S_{\rm DBI} = S_0 + S_{\rm YM} + \CO(\CF^3) . 
$$ 
In a similar way to the anti-podal case it is convenient 
to introduce a new coordinate $z$ defined by\foot{
The $\zeta$ parameter is a measure of the ``string endpoint mass'' 
of the quark. The latter is defined as 
$$
m^s_q= {1 \over 2\pi\alpha'}\int_{u_{\rm KK}}^{u_0} du \sqrt{g_{00}g_{uu}} . 
$$
This quantity is neither the QCD mass nor the constituent mass of the quark. 
In a crude way a non-spinning meson has a mass of 
the form $M= T_{\rm st} L + 2m^s_q$ (for equal endpoints).
}
\eqn\zcoord{ 
u=u_{\rm KK}(\zeta^3 + 
\zeta z^2)^{1 \over 3} ,\quad \zeta = {u_0 \over u_{\rm KK}}. 
} 
$z$ and $\zeta$ are dimensionless. 
$\zeta$ takes a value in $[1,\infty)$ 
because of $u_0 \geq u_{\rm KK}$, 
while $z$ takes a value in $(-\infty,\infty)$. 
Though \uxfour\ implies that $x_4(u)$ is a double-valued function, 
the $z$ coordinate makes it single-valued. 
The Yang-Mills part $S_{\rm YM}$ is calculated 
in terms of \DeightIndmet\ and \zcoord, 
\eqn\YMorig{ 
S_{\rm YM} = -\kappa \int 
d^4xdz\, \Tr\biggl[{1 \over 2}h(z;\zeta)\CF_{\mu\nu}^2 +M_{\rm KK}^2 
k(z;\zeta)\CF_{\mu z}^2 \biggr] , 
} 
where 
$$\eqalign{ 
h(z;\zeta) &= \sqrt{\zeta^2 z^2 (\zeta^3+\zeta z^2) \over 
(\zeta^3+\zeta z^2)^{8 \over 3}-(\zeta^3+\zeta z^2)^{5 \over 3} 
-\zeta^8 +\zeta^5} , \cr 
k(z;\zeta) &= (\zeta^3+\zeta z^2)^{1 \over 
6} \sqrt{(\zeta^3+\zeta z^2)^{8 \over 3}-(\zeta^3+\zeta z^2)^{5 
\over 3} -\zeta^8 +\zeta^5 \over \zeta^2 z^2} 
}$$ 
and $\kappa := \lambda N_c/(216\pi^3)$. 
$\lambda$ is t'Hooft coupling, $\lambda := g_{\rm YM}^2 N_c$. 
It is easy to check that 
for $\zeta=1$ the anti-podal case is reproduced, 
namely, $h(z)= (1+z^2)^{-1/3}$ and $k(z)= 1+z^2$. 

From now on, we use the $M_{\rm KK}=1$ unit. 
When necessary later, 
we shall be able to easily recover the factor $M_{\rm KK}$. 
For the later convenience, we rescale the coordinate $z$ 
and the field $\CA_z$, 
\eqn\tilz{ 
\Tz := \sqrt{h_0 \over k_0}z ,\quad \CA_\Tz := 
\sqrt{k_0 \over h_0}\CA_z . 
} 
$h_0$ and $k_0$ are defined through the expansions 
$h(z;\zeta) = h_0(\zeta) + \CO(z^2)$ and 
$k(z;\zeta) = k_0(\zeta) + \CO(z^2)$ respectively, 
\eqn\hkzero{ 
h_0 = \zeta 
\sqrt{3 \over 8\zeta^3-5} ,\quad 
k_0 = \zeta \sqrt{8\zeta^3-5 \over 3} . 
} 
It is clear from these expressions that there is a critical 
value of $\zeta$, that is $\zeta_{cr}= (5/8)^{1/3}(<1)$, 
such that necessarily $\zeta>\zeta_{cr}$. 
Recall however that by its definition $\zeta \geq 1$. 
We will come back to this point at the end of this section. 

The action \YMorig\ is rewritten as 
\eqn\YMact{ 
S_{\rm YM} = 
-{\tilde \kappa}(\zeta) \int d^4xd\Tz\, \Tr\biggl[{1 \over 
2}\Th(\Tz;\zeta)\CF_{\mu\nu}^2 +\Tk(\Tz;\zeta)\CF_{\mu \Tz}^2 
\biggr]  , 
} 
where $\Th(\Tz;\zeta)$ and $\Tk(\Tz;\zeta)$ are defined 
in terms of \tilz\ and \hkzero\ by 
\eqn\DefThk{ 
\Th(\Tz;\zeta) := {h(z;\zeta) \over h_0(\zeta)} , \quad 
\Tk(\Tz;\zeta) := {k(z;\zeta) \over k_0(\zeta)} , \quad 
{\tilde \kappa}(\zeta) := \kappa\sqrt{h_0(\zeta) k_0(\zeta)}= \kappa\zeta . 
} 
Since $\Th(0;\zeta) = \Tk(0;\zeta) = 1$, $\Th(\Tz,\zeta)$ and 
$\Tk(\Tz,\zeta)$ can be expanded with respect to $\Tz$ as 
\eqn\expThk{ 
\Th(\Tz;\zeta) = 1 + \sum_{n=1}^\infty \Th_n(\zeta)\Tz^{2n} ,\quad 
\Tk(\Tz;\zeta) = 1 + \sum_{n=1}^\infty \Tk_n(\zeta)\Tz^{2n} . 
} 
For example, $\Th_1(\zeta)$ and $\Tk_1(\zeta)$ are evaluated 
\eqn\Thkone{ 
\Th_1(\zeta) = {2\zeta^3-5 \over 9\zeta^2} ,\quad \Tk_1(\zeta) = 
{14\zeta^3-5 \over 9\zeta^2} . 
} 
 
We shall concentrate on the simplest non-ablelian $N_f=2$ case. 
With the final goal of comparing the theoretical results to 
the experimental data of baryons, 
it makes sense to choose this case, 
since the up and down quarks have almost the same mass 
and are much lighter than a strange quark. 
The $U(2)$ gauge field is decomposed, 
\eqn\decgauge{ 
\CA = A + {1 \over \sqrt{2N_f}}\HA = A + {1 \over 2}\HA, 
} 
where $A$ and $\HA$ denote the $SU(2)$ and $U(1)$ gauge fields respectively. 
The Chern-Simon action \NfCSact\ with the rescaling \tilz\ is 
written down as 
\eqn\CSact{ 
S_{\rm CS} = {27\pi\kappa \over 
8\lambda}\epsilon^{\alpha_1\alpha_2\alpha_3\alpha_4\alpha_5} \int 
d^4x d\Tz \biggl[ 
    \HA_{\alpha_1}\tr(F_{\alpha_2\alpha_3}F_{\alpha_4\alpha_5}) 
    +{1 \over6}\HA_{\alpha_1}\HF_{\alpha_2\alpha_3}\HF_{\alpha_4\alpha_5}  
	\biggr] 
} 
up to total derivatives. 
The indices $\alpha_i$ are $0,1,2,3,\Tz$ and $\epsilon^{0123\Tz}=1$. 
 
The action of the gauge fields considered in this paper 
is constructed from \YMact\ and \CSact, 
\eqn\gaugeact{ 
S_{\rm gauge} = S_{\rm YM} + S_{\rm CS} . 
} 
This action leads to the following equations of motion for the gauge fields: 
\eqna\EOMgauge 
$$\eqalignno{ 
&\Th(\Tz)D_\nu F^{\mu\nu} +D_\Tz \bigl( \Tk(\Tz) F^{\mu\Tz} \bigr) 
    = {27\pi\kappa \over 8\lambda\Tkappa} 
    \epsilon^{\mu\alpha_1\alpha_2\alpha_3\alpha_4} 
    \HF_{\alpha_1\alpha_2}F_{\alpha_3\alpha_4} , &\EOMgauge{a} \cr 
&\Tk(\Tz) D_\mu F^{\Tz\mu} 
    = {27\pi\kappa \over 8\lambda\Tkappa} 
    \epsilon^{\Tz\mu_1\mu_2\mu_3\mu_4} 
    \HF_{\mu_1\mu_2}F_{\mu_3\mu_4} , &\EOMgauge{b} \cr 
&\Th(\Tz)\partial_\nu \HF^{\mu\nu} +\partial_\Tz \bigl( \Tk(\Tz) \HF^{\mu\Tz} \bigr) 
    = {27\pi\kappa \over 8\lambda\Tkappa} 
    \epsilon^{\mu\alpha_1\alpha_2\alpha_3\alpha_4} 
    \biggl[ \tr (F_{\alpha_1\alpha_2}F_{\alpha_3\alpha_4}) 
    +{1 \over 2}\HF_{\alpha_1\alpha_2}\HF_{\alpha_3\alpha_4} \biggr] , \cr 
&&\EOMgauge{c} \cr 
&\Tk(\Tz) \partial_\mu \HF^{\Tz\mu} 
    = {27\pi\kappa \over 8\lambda\Tkappa} 
    \epsilon^{\Tz\mu_1\mu_2\mu_3\mu_4} 
    \biggl[ \tr (F_{\mu_1\mu_2}F_{\mu_3\mu_4}) 
    +{1 \over 2}\HF_{\mu_1\mu_2}\HF_{\mu_3\mu_4} \biggr] , &\EOMgauge{d} 
}$$ 
where $\mu_i,\nu$ are $0,1,2,3$.

%%%%%%%%%%%%%%%%%%%%%%%%%%%%%% 
\subsec{Baryon as instanton} 
 
Following \HSSY, we now introduce the rescaling 
of the coordinates and the fields, 
\eqn\lamscale{\eqalign{ 
&x^0 = 
x_{(r)}^0,\quad x^i = {1 \over \sqrt{\lambda}}x_{(r)}^i,\quad \Tz = 
{1 \over \sqrt{\lambda}}\Tz_{(r)} , \cr 
&\CA_0 = \CA_{(r)0},\quad 
\CA_i = \sqrt{\lambda}\CA_{(r)i},\quad \CA_\Tz = 
\sqrt{\lambda}\CA_{(r)\Tz} , 
}} 
where $i=1,2,3$, and consider the expansion 
with respect to large $\lambda$. 
Under this expansion, we can approximate 
$\Th(\Tz_{(r)}/\sqrt{\lambda};\zeta) \approx 1$ and 
$\Tk(\Tz_{(r)}/\sqrt{\lambda};\zeta) \approx 1$ from \expThk. 
The equations of motion \EOMgauge{} are then reduced 
at the leading order of $\lambda$ to 
\eqna\EOMglam 
$$\eqalignno{ 
D^{(r)}_M F_{(r)}^{NM} &= 0 , &\EOMglam{a} \cr 
D^{(r)}_M F_{(r)}^{0M} &= {27\pi\kappa \over 8\Tkappa}\epsilon_{MNPQ}\HF_{(r)}^{MN}F_{(r)}^{PQ} , &\EOMglam{b} \cr 
\partial^{(r)}_M \HF_{(r)}^{NM} &= 0 , &\EOMglam{c} \cr 
\partial^{(r)}_M \HF_{(r)}^{0M} &= {27\pi\kappa \over 8\Tkappa}\epsilon_{MNPQ} 
        \biggl[ \tr \Bigl(F_{(r)}^{MN}F_{(r)}^{PQ}\Bigr) +{1 \over 2}\HF_{(r)}^{MN}\HF_{(r)}^{PQ} \biggr] , &\EOMglam{d} 
}$$ 
where $M,N,P,Q = 1,2,3,\Tz$. 
Since \EOMglam{a} is a four-dimensional instanton equation, 
its classical solution can be described as BPST instanton \BPST, 
\eqnn\instsol 
$$\eqalignno{ 
A^{\rm cl}_M(x^i,\Tz) &\Bigl(= \sqrt{\lambda} A_{(r)M}(x_{(r)}^i,\Tz_{(r)}) \Bigr) 
= -i v(\xi)g\partial_M g^{-1} &\instsol \cr 
&v(\xi) = {\xi^2 \over \xi^2 + \rho^2} ,\quad 
\xi = \sqrt{(x^i - X^i)^2 + (\Tz-\TZ)^2} , \cr 
&g(x^i,z) = {(\Tz-\TZ){\bf 1} -i(x^i-X^i)\tau_i \over \xi}. \quad (i=1,2,3) 
}$$ 
The field strength of $A^{\rm cl}_M$ are calculated as 
$$\eqalign{ 
F^{\rm cl}_{ij} = {2 \rho^2 \over (\xi^2+\rho^2)^2}\epsilon^{ija}\tau_a ,\quad 
F^{\rm cl}_{\Tz j} = {2 \rho^2 \over (\xi^2+\rho^2)^2}\tau_j . 
}$$ 
This solution is a one-instanton solution. 
In a similar way we can write a 't Hooft multi-instanton solution. 
The equations \EOMglam{b,c} lead to 
\eqn\aHfsol{ 
A^{\rm cl}_0 = \HA^{\rm cl}_M = 0 
} 
with an appropriate gauge fixing. 
Substituting the solutions \instsol\ and \aHfsol\ into 
the equation of motion for $\HA_0$ \EOMglam{d}, 
we obtain 
$$ 
\partial_M^2\HA_0 = -{648\pi \kappa \over \lambda \Tkappa}{\rho^4 \over (\xi^2+\rho^2)^4} , 
$$ 
which can be solved, 
\eqn\Hazerosol{ 
\HA^{\rm cl}_0 = {27\pi\kappa \over \lambda\Tkappa}{\xi^2 +2\rho^2 \over (\xi^2 + \rho^2)^2} . 
} 
Here we should note that the $\zeta$ dependence is included 
in the factor $\kappa/\Tkappa = \zeta^{-1}$. 
This factor does not appear 
in the other gauge fields $A_0,A_M,\HA_M$ 
and these fields are in the order of $\lambda^0$. 
On the other hand, $\HA_0$ is in the order of $\lambda^{-1}$, 
that is to say, the $\zeta$ dependence is derived 
from the $\lambda^{-1}$ correction. 
 
In terms of the classical solutions \instsol, \aHfsol\ and 
\Hazerosol, 
one can compute the mass of the baryon $M$, 
which depends on the moduli parameters $\rho,Z$ 
via $S=-\int dt\, M$ from the action 
\gaugeact, 
$$\eqalign{ 
M &= 8\pi^2\Tkappa \Biggl[1 +{\Th_1+\Tk_1 \over 
2}\biggl(\TZ^2+{\rho^2 \over 2}\biggr) +\biggl({27\pi\kappa \over 
\lambda\Tkappa}\biggr)^2{1 \over 5\rho^2} \Biggr]\cr 
&= 8\pi^2\kappa\zeta\biggl(1 +{1 \over 3\zeta^2}Z^2 +{8\zeta^3-5 \over 
18\zeta^2}\rho^2 +{729\pi^2 \over 5\lambda^2\zeta^2}{1 \over 
\rho^2}\biggr) , \cr 
}$$ 
where we used \Thkone. 
Then we can find the critical values of the moduli parameters 
so that $M$ is minimized, 
\eqn\barcri{ 
Z_{\rm cr} = 0 ,\quad \rho_{\rm cr}^2 = {81\pi \over 
\lambda}\sqrt{2 \over 40\zeta^3-25} , 
} 
and the minimum value of $M$ becomes 
$$ 
M_{\rm min} = 8\pi^2\kappa \Biggl( \zeta + {18\pi \over 
\lambda\zeta}\sqrt{8\zeta^3-5 \over 10} \Biggr) . 
$$ 
 
From the expression of $\rho_{\rm cr}$ we thus see that 
generalizing the anit-podal case to the $\zeta\geq 1$ family of models 
does not improve the situation 
that the size of the baryon scales like $\sim 1/\sqrt{\lambda}$ 
and hence stringy corrections can play a role in the game. 
 
The same kind of analysis can be done in the non-critical 
holographic model in six dimensions \refs{\KSi,\KSii}. 
This will be discussed in Section 6.

%%%%%%%%%%%%%%%%%%%%%%%%%%%%%% 
\subsec{Mass spectra} 
The study of the mass spectra of the baryons 
is also very similar to the one in \HSSY. 
The idea is to introduce the collective coordinates 
associated with the instanton solution 
and to semi-classically quantize them. 
The collective coordinates of instanton span 
a moduli space with a topology of 
$\Bbb{R}^4 \times (\Bbb{R}^4/\Bbb{Z}_2)$. 
The moduli are the position $(X^i,Z)$, 
the size $\rho = \sqrt{y_1^2 +\cdots+ y_4^2}$ 
and the $SU(2)$ orientation $a_I := y_I/\rho$ $(I=1,\dots,4)$. 
As usual the basic assumption of the semi-classical quantization is 
that the collective coordinates $X^\alpha := (X^i,Z,y_I)$ 
depend on time. 

Thus the  fluctuations of $SU(2)$ gauge fields are described as 
$$ 
A_M(t,x) = V(t,x^i) A_M^{\rm cl}(x^i,z;X^\alpha(t))V^{-1}(t,x^i) 
    -iV(t,x^i) \partial_M V^{-1}(t,x^i) , 
$$ 
where $A_M^{\rm cl}$ has been given by \instsol. 
The equation of motion \EOMglam{b} determines $\Phi := -iV^{-1}{\dot V}$ as 
\eqnn\Phidecomp 
$$\eqalignno{ 
\Phi(t,x) &= -{\dot X}^i(t) A^{\rm cl}_i(x) -{\dot \TZ}(t)A^{\rm cl}_\Tz(x) + \chi^a(t)\Phi_a(x) , \cr 
&\chi^a = 2(a_4{\dot a}_a -{\dot a}_4 a_a +\epsilon^{abc}a_b{\dot a}_c) , \quad 
\Phi_a = {1 \over 2}v(\xi) g \tau_a g^{-1} .  &\Phidecomp 
}$$ 
In terms of these equations, the field strength of the $SU(2)$ gauge field is 
written as $F_{MN} = V F^{\rm cl}_{MN} V^{-1}$ and 
$F_{0M} = V ({\dot X}^N F^{\rm cl}_{MN} +{\dot \rho}\partial_\rho A^{\rm cl}_M -\chi^a D^{\rm cl}_M\Phi_a)V^{-1}$. 
The equation of motion \EOMglam{d} with this solution of $A_M$ 
does not change $\HA_0$, that is, $\HA_0 = \HA^{\rm cl}_0$. 
 
Substituting the gauge fields obtained so far into the action 
\YMact, derives a  Lagrangian of the collective coordinates which is 
the same as in \HSSY, 
\eqn\collLag{ 
L = -m_0 + {1 \over 2}m_X {\dot 
\vecX}^2 +{1 \over 2}m_Z {\dot Z}^2 -{1 \over 2}m_Z\omega_Z^2 Z^2 
+{1 \over 2}m_y {\dot \vecy}^2 -{1 \over 2}m_y\omega_\rho^2 \rho^2 
    -{Q \over \rho^2} , 
} 
where ${\dot \vecy}^2 = {\dot \rho}^2 +\rho^2 {\dot \veca}^2$ 
apart from the fact that the various mass parameters are now $\zeta$ 
dependent as follows 
\eqna\coefcoll 
$$\eqalignno{ 
&m_0 = m_X = 8\pi^2\Tkappa = 8\pi^2\kappa\zeta , &\coefcoll{a} \cr 
&m_Z = 8\pi^2\kappa {3\zeta \over 8\zeta^3-5} ,\quad \omega_Z^2 = 
{16\zeta^3 -10 \over 9\zeta^2} , &\coefcoll{b} \cr &m_y = 
16\pi^2\kappa \zeta ,\quad \omega_\rho^2 = {8\zeta^3 -5 \over 
18\zeta^2} ,\quad Q = 8\pi^2\kappa{729 \pi^2 \over 5\lambda^2\zeta} 
. &\coefcoll{c} 
}$$ 
The system is then quantized in the same way as \HSSY. 
Using the canonical momenta, the corresponding Hamiltonian becomes 
$H = -(2m_0)^{-1}(\partial/\partial \vecX)^2 
-(2m_0)^{-1}(\partial/\partial \TZ)^2 -(4m_0)^{-1}(\partial/\partial 
\vecy)^2 +U$. 
The isospin and spin currents are defined by 
\eqnn\isosope 
\eqnn\spinope 
$$\eqalignno{ 
I_a &= {i \over 2}\biggl(y_4{\partial \over \partial y_a} 
-y_a{\partial \over \partial y_4} -\epsilon_{abc}y_b{\partial \over 
\partial y_c}\biggr) , &\isosope \cr 
J_a &= {i \over 2}\biggl(-y_4{\partial \over \partial y_a} +y_a{\partial \over 
\partial y_4} -\epsilon_{abc}y_b{\partial \over \partial y_c}\biggr) 
. &\spinope 
}$$ 
For a  baryon which  is located at $\vecX = 0$, in other words, 
the baryon is static with respect to fluctuations in 
the ordinary four-dimensional spacetime. 
The energy spectra of the fluctuations of $Z$ and $\vecy$ 
take the following form 
\eqn\collspec{ 
E_y = \omega_\rho \Bigl(\sqrt{(l+1)^2 +2 m_y Q} + 
2n_\rho +1 \Bigr) , \quad E_Z = \omega_Z\biggl(n_z +{1 \over 
2}\biggr) , 
} 
and hence, using \coefcoll{}, the baryon mass formula is given by  
\eqnn\baryonmass 
$$\eqalignno{ 
M_{l,n_\rho,n_z} &= m_0 + E_y + E_Z \cr 
&= 8\pi^2\kappa\zeta 
    +\sqrt{8\zeta^3-5 \over 3\zeta^2} \Biggl[\sqrt{{(l+1)^2 \over 6} +{2N_c^2 \over 15}} +{2(n_\rho + n_z) +2 \over \sqrt{6}}\Biggr] . &\baryonmass 
}$$ 
$l$ is a positive odd integer and describes a spin $J$ 
and an isospin $I$ as $I=J=l/2$. 
For later convenience, we write down the wave functions 
of proton $|p\uparrow\rangle$ and neutron $|n\uparrow\rangle$, 
\eqnn\pnWF 
\eqna\WFrhoZ 
$$\eqalignno{ 
&|p\uparrow\rangle \propto R(\rho;\zeta)\psi_Z(Z;\zeta) (a_1 +ia_2) ,\quad 
|n\uparrow\rangle \propto R(\rho;\zeta)\psi_Z(Z;\zeta) (a_4 +ia_3) , &\pnWF \cr 
&R(\rho;\zeta) = \rho^{-1+2\sqrt{1+N_c^2/5}} 
    \exp\Biggl(-m_0\sqrt{8\zeta^3-5 \over 18\zeta^2}\rho^2\Biggr) , &\WFrhoZ{a}\cr 
&\psi_Z(Z;\zeta) = \exp\Biggl(-{m_0 \over \sqrt{2\zeta^2(8\zeta^3-5)}} Z^2\Biggr) . &\WFrhoZ{b} 
}$$ 

At this point we would like to compare the baryon masses 
and in particular the mass differences between the various baryonic states. 
For this purpose we first have to turn on back $M_{\rm KK}$. 
If we identify the modes of $(l,n_\rho,n_z) = (1,0,0)$ and $(3,0,0)$ with 
$n(940)$ and $\Delta(1232)$ (see also Table 1), 
$\zeta$ and $M_{\rm KK}$ satisfy 
\eqnn\barNoo \eqnn\barDoo 
$$\eqalignno{ 
&{N_c \lambda \over 27\pi}\zeta 
    +\sqrt{8\zeta^3-5 \over 3\zeta^2}\Biggl(\sqrt{{2 \over 3} +{6 \over 5}} +\sqrt{2 \over 3} \Biggr) 
    = {940 \over M_{\rm KK}} , &\barNoo \cr 
&{N_c \lambda \over 27\pi}\zeta 
    +\sqrt{8\zeta^3-5 \over 3\zeta^2}\Biggl(\sqrt{{8 \over 3} +{6 \over 5}} +\sqrt{2 \over 3} \Biggr) 
    = {1232 \over M_{\rm KK}} . &\barDoo 
}$$ 
We can read from these equations, 
\eqn\estI{ 
M_{\rm KK} \sqrt{8\zeta^3-5 \over 3\zeta^2} 
    = {292\sqrt{15} \over \sqrt{58} -\sqrt{28}} . 
} 
Since the left hand side of this equation is 
the monotonically increasing function of $\zeta$, 
the Kaluza-Klein mass $M_{\rm KK}$ is bounded as 
\eqn\KKbound{ 
M_{\rm KK} \leq {292\sqrt{15} \over \sqrt{58} -\sqrt{28}} 
    \approx 487 \,{\rm [MeV]}. 
} 
In terms of \barNoo\ (or \barDoo) and \estI, 
we can now compute the baryon masses $M_{\rm KK} M_{l,n_\rho,n_z}$[MeV], 
which are shown in Table 2 and compare them 
to the experimental data of Table 1. 
This is done by first fixing $N_c = 3$. 
\bigskip 
\centerline{\vbox{\offinterlineskip \halign{ \hskip2pt #\strut\hfil 
\hskip1pc & \hfil#\strut\hfil &\hskip1pt \vrule# \hskip3pt & 
#\strut\hfil \hskip1pc & \hfil#\strut\hfil \hskip1pt \cr 
\noalign{\hrule} $N$ baryons & $I\bigl(J^P\bigr)$ && $\Delta$ 
baryons & $I\bigl(J^P\bigr)$ \cr \noalign{\hrule \vskip1pt \hrule} 
$n(940)$ & ${1\over 2}\bigl({1 \over 2}^+\bigr)$ && $\Delta(1232)$ & 
${3\over 2}\bigl({3 \over 2}^+\bigr)$ \cr $N(1440)$ & ${1\over 
2}\bigl({1 \over 2}^+\bigr)$ && $\Delta(1600)$ & ${3\over 2}\bigl({3 
\over 2}^+\bigr)$ \cr $N(1535)$ & ${1\over 2}\bigl({1 \over 
2}^-\bigr)$ && $\Delta(1700)$ & ${3\over 2}\bigl({3 \over 
2}^-\bigr)$\cr $N(1650)$ & ${1\over 2}\bigl({1 \over 2}^-\bigr)$ && 
$\Delta$(1920) & ${3\over 2}\bigl({3 \over 2}^+\bigr)$\cr $N(1710)$ 
& ${1\over 2}\bigl({1 \over 2}^+\bigr)$ && $\Delta(1940)$ & ${3\over 
2}\bigl({3 \over 2}^-\bigr)$\cr $N(2090)$ & ${1\over 2}\bigl({1 
\over 2}^-\bigr)$ && & \cr $N(2100)$ & ${1\over 2}\bigl({1 \over 
2}^+\bigr)$ && & \cr \noalign{\hrule} } }}\nobreak \centerline{Table 
1: The experimental data of baryon mass spectra \PDG. } 
\bigskip\noindent
\bigskip
\centerline{\vbox{\offinterlineskip 
\halign{ 
\hskip2pt #\strut\hfil \hskip1pc 
& \hfil#\strut\hfil \hskip1pc 
& #\strut\hfil 
&\hskip2pt \vrule# \hskip3pt 
& #\strut\hfil \hskip1pc 
& \hfil#\strut\hfil \hskip1pc 
& #\strut\hfil \hskip1pt \cr 
\noalign{\hrule} 
$N$ baryons & $(n_\rho, n_z)$ & $M_{\rm KK} M_{1,n_\rho,n_z}$ && $\Delta$ baryons & $(n_\rho, n_z)$ & $M_{\rm KK} M_{3,n_\rho,n_z}$ \cr 
\noalign{\hrule \vskip1pt \hrule} 
$n(940)$ & $(0,0)$ & 940 && 
$\Delta(1232)$ & $(0,0)$ & 1232 \cr 
$N(1440)$ & $(1,0)$ & 1337 && 
$\Delta(1600)$ & $(1,0)$ & 1629 \cr 
$N(1535)$ & $(0,1)$ & 1337 && 
$\Delta(1700)$ & $(0,1)$ & 1629 \cr 
$N(1650)$ & $(1,1)$ & 1735 && 
$\Delta$(1920) & $(2,0),(0,2)$ & 2027 \cr 
$N(1710)$ & $(2,0),(0,2)$ & 1735 && 
$\Delta(1940)$ & $(1,1)$ & 2027 \cr 
$N(2090)$ & $(2,1),(0,3)$ & 2132 && 
& \cr 
$N(2100)$ & $(1,2),(3,0)$ & 2132 && 
& \cr 
\noalign{\hrule} 
} 
}}\nobreak 
\centerline{Table 2: The baryon mass spectra in our model.} 
\bigskip\noindent 
Since the $1/N_c$ corrections are important 
for the states of larger quantum numbers, 
it is physically better to fit the baryon mass formula \baryonmass\ 
to the experimental data by using the lower quantum numbers. 
But here instead we determine the masses by using a best fit approach, 
namely, minimizing  $\chi^2$  with respect to the all states listed 
in Table 1. 
We need to determine the two parameters $A$ and $B$ 
which are defined from \baryonmass\ by 
$$\eqalign{ 
M_{\rm KK} M_{l,n_\rho,n_z} 
&= A + B \Biggl[\sqrt{{(l+1)^2 \over 6} +{6 \over 5}} +{2(n_\rho + n_z) +2 \over \sqrt{6}}\Biggr] , \cr 
&A:= {M_{\rm KK} \lambda \over 9\pi}\zeta ,\quad 
B := M_{\rm KK} \sqrt{8\zeta^3-5 \over 3\zeta^2} . 
}$$ 
Though we should take care of the zero point energy, 
here it can be absorbed into $A$. 
Then $(A,B)=(99.9,424.8)$ is the best fit. 
This implies that $\lambda$ is not large and 
hence $1/\lambda$ corrections may not be negligable.
The Kaluza-Klein mass is bounded so that $M_{\rm KK} \leq 424.8$ [MeV]. 
The mass spectra evaluated in terms of these values are shown in Table 3. 
\bigskip 
\centerline{\vbox{\offinterlineskip 
\halign{ 
\hskip2pt #\strut\hfil \hskip1pc 
& \hfil#\strut\hfil \hskip1pc 
& #\strut\hfil 
&\hskip2pt \vrule# \hskip3pt 
& #\strut\hfil \hskip1pc 
& \hfil#\strut\hfil \hskip1pc 
& #\strut\hfil \hskip1pt \cr 
\noalign{\hrule} 
$N$ baryons & $(n_\rho, n_z)$ & $M_{\rm KK} M_{1,n_\rho,n_z}$ && $\Delta$ baryons & $(n_\rho, n_z)$ & $M_{\rm KK} M_{3,n_\rho,n_z}$ \cr 
\noalign{\hrule \vskip1pt \hrule} 
$n(940)$ & $(0,0)$ & 1027 && 
$\Delta(1232)$ & $(0,0)$ & 1282 \cr 
$N(1440)$ & $(1,0)$ & 1374 && 
$\Delta(1600)$ & $(1,0)$ & 1629 \cr 
$N(1535)$ & $(0,1)$ & 1374 && 
$\Delta(1700)$ & $(0,1)$ & 1629 \cr 
$N(1650)$ & $(1,1)$ & 1721 && 
$\Delta$(1920) & $(2,0),(0,2)$ & 1976 \cr 
$N(1710)$ & $(2,0),(0,2)$ & 1721 && 
$\Delta(1940)$ & $(1,1)$ & 1976 \cr 
$N(2090)$ & $(2,1),(0,3)$ & 2068 && 
& \cr 
$N(2100)$ & $(1,2),(3,0)$ & 2068 && 
& \cr 
\noalign{\hrule} 
} 
}}\nobreak 
\centerline{Table 3: The baryon masses by the use of the minimal $\chi^2$ fitting.} 
\bigskip\noindent 
Since there are more degeneracies for the states with larger quantum numbers, 
the $\chi^2$-fitted data are strongly affected by these states.

%%%%%%%%%%%%%%%%%%%%%%%%%%%%%% 
\subsec{Mean radii, magnetic moments and couplings} 

Next we should like to determine 
the impact of $\zeta \neq 1$ on the baryonic properties 
of the mean radii, magnetic moments and various couplings. 
For that purpose we consider the currents 
of the $U(N_f)_L \times U(N_f)_R$ chiral symmetry 
in the same way as was done in \HSS.  
On account of the gauge configuration 
$$\eqalign{ 
&\CA_\alpha(x^\mu, \Tz) = \CA^{\rm cl}_\alpha(x^\mu, \Tz) +\delta\CA_\alpha(x^\mu, \Tz) , \cr 
&\delta\CA_\alpha(x^\mu, +\infty) = \CA_{L\mu}(x^\mu) ,\quad 
\delta\CA_\alpha(x^\mu, -\infty) = \CA_{R\mu}(x^\mu) , 
}$$ 
we can read the currents from the action \gaugeact, 
$$ 
S_{\rm gauge} = -2 \int d^4x\, \Tr\bigl(A_{L\mu}\CJ^\mu_L +A_{R\mu}\CJ^\mu_R \bigr) +\CO(\delta\CA^2 ) , 
$$ 
where 
\eqn\LRcur{ 
\CJ_{L\mu} = -\Tkappa \Bigl(\Tk(\Tz)\CF^{\rm 
cl}_{\mu\Tz}\Bigr)\Big|_{\Tz = +\infty} ,\quad \CJ_{R\mu} = \Tkappa 
\Bigl(\Tk(\Tz)\CF^{\rm cl}_{\mu\Tz}\Bigr)\Big|_{\Tz = -\infty} . 
} 
Obviously from the left and right currents one can form 
the vector and axial currents as follows, 
\eqnn\Vcur 
\eqnn\Acur 
$$\eqalignno{ 
\CJ_{V\mu} &= \CL_{L\mu} + \CJ_{R\mu} 
= -\Tkappa\Bigl[\Tk(\Tz)\CF^{\rm cl}_{\mu\Tz}\Bigr]^{\Tz=+\infty}_{\Tz=-\infty} , &\Vcur \cr 
\CJ_{A\mu} &= \CL_{L\mu} - \CJ_{R\mu} 
= -\Tkappa\Bigl[\Tk(\Tz)\CF^{\rm cl}_{\mu\Tz} \psi_0(\Tz)\Bigr]^{\Tz=+\infty}_{\Tz=-\infty} . &\Acur 
}$$ 
$\psi_0(\Tz)$ is defined by $\psi_0(\Tz) := \xi(\Tz)/\xi(\infty)$ 
in terms of the function $\xi(\Tz)$ satisfying the equation 
$\Tk(\Tz)\partial_\Tz\xi(\Tz)=1$. 
$\xi(\Tz)$ can be rewritten as 
\eqn\defxi{ 
\xi(\Tz) = \int_0^\Tz {d\Tz' \over \Tk(\Tz')} , 
} 
because $\xi(\Tz)$ is an odd function and $\xi(0) = 0$. 
Then $\psi_0(\Tz)$ has the property of $\psi_0(\pm\infty) = \pm 1$. 
The currents are also decomposed as the gauge fields \decgauge\ to 
the $SU(2)$ and $U(1)$ parts, $\CJ^\mu = J^\mu + (1/2)\HJ^\mu$. 
In order to evaluate the currents \Vcur\ and \Acur, 
it is necessary to understand 
the behavior of the gauge field strengths 
at the UV boundary, $\Tz = \pm \infty$. 
But so far we know the expression of the gauge field strengths 
only in the region of $\Tz \ll 1$. 
Ref.\HSS\ has succeeded in extending it to the large $\Tz$ region in 
the anti-podal case $(\zeta =1)$. 
In the same way, we can easily evaluate in the non-anti-podal case 
the gauge field strengths for $\TZ \ll 1 \ll \Tz$, 
\eqna\GFS 
$$\eqalignno{ 
F_{0\Tz} &\approx 2\pi^2\partial_0\bigl(\rho^2 {\bf a}\tau^a{\bf a}^{-1}\bigr)\partial_a H 
    -4\pi^2i\rho^2{\bf a}{\dot {\bf a}}^{-1}\partial_\Tz G \cr 
&\quad  -2\pi^2\rho^2{\bf a}\tau^a{\bf a}^{-1}{\dot X}^i \bigl\{ \bigl(\partial_i\partial_a -\delta_{ia}\partial_j^2 \bigr)H -\epsilon_{iaj}\partial^j\partial_\Tz G \bigr\} , &\GFS{a} \cr 
F_{i\Tz} &\approx 2\pi^2\rho^2{\bf a}\tau^a{\bf a}^{-1} \bigl\{ \bigl(\partial_i\partial_a -\delta_{ia}\partial_j^2 \bigr)H -\epsilon_{iaj}\partial^j\partial_\Tz G \bigr\} , &\GFS{b} \cr 
\HF_{0\Tz} &\approx {108\pi^3\kappa \over \lambda\Tkappa}\partial_\Tz G , &\GFS{c} \cr 
\HF_{i\Tz} &\approx {108\pi^3\kappa \over \lambda\Tkappa} 
    \biggl[{\dot \TZ} \partial_iH -{\dot X}_i\partial_\Tz G 
    -{\rho^2\chi^a \over 4}\bigl\{ \bigl(\partial_i\partial_a -\delta_{ia}\partial_j^2 \bigr)H -\epsilon_{iaj}\partial^j\partial_\Tz G \bigr\}\biggr] , &\GFS{d} 
}$$ 
where ${\bf a}=a_4+ia_a\tau^a$. 
$H$ and $G$ are the Green's functions generalised for the curved background, 
\eqn\Grefunc{ 
G = \Tkappa \sum_{n=1}^\infty \psi_n(\Tz)\psi_n(\TZ)Y_n(|\vecx -\vecX|) ,\quad 
H = \Tkappa \sum_{n=0}^\infty \phi_n(\Tz)\phi_n(\TZ)Y_n(|\vecx -\vecX|) . 
} 
The eigen functions $\psi_n$'s are defined by 
\eqn\EFphi{ 
-\Th(\Tz)^{-1}\partial_\Tz \bigl(\Tk(\Tz)\partial_\Tz\psi_n \bigr) = \lambda_n\psi_n ,\quad 
\Tkappa \int d\Tz\, \Th(\Tz)\psi_m\psi_n = \delta_{mn} , 
} 
while $\phi_n$'s are defined on account of \defxi\ by 
\eqn\EFpsi{ 
\phi_0(\Tz) = {1 \over \sqrt{2\Tkappa \xi(\infty)}\Tk(\Tz)} , \quad 
\phi_n(\Tz) = {1 \over \sqrt{\lambda_n}}\partial_\Tz\psi_n(\Tz) , \quad (n \in \Bbb{N}) 
} 
so that these modes satisfy the normalisation 
$\Tkappa \int d\Tz\, \Tk(\Tz)\phi_m\phi_n = \delta_{mn}$ 
for $n,m \in \{0,\Bbb{N}\}$. 
$Y_n$ denotes the Yukawa potential\foot{ 
The eigen equation in \EFphi\ is rewritten through \tilz\ and \DefThk\ as 
$-h(z)^{-1}\partial_z \bigl(k(z)\partial_z\psi_n \bigr) = \lambda_n\psi_n$, 
which is exactly the eigen equation providing the meson mass spectra. 
That is to say, the meson mass $m_n$ is denoted by $m_n = \sqrt{\lambda_n}$. 
} 
\eqn\Yukawapot{ 
Y_n(r) = -{1 \over 4\pi}{e^{-\sqrt{\lambda_n}r} \over r} . 
} 
 
\medskip 
\noindent$\underline{\hbox{\it Mean square radii}}$ 
\medskip 
The baryon number current is denoted in terms of the vector current \Vcur\ by 
\eqn\BNcur{ 
J_B^\mu = {2 \over N_c}\HJ_V^\mu 
    = -{2 \over N_c}\Tkappa \Bigl[\Tk(\Tz)\HF^{\mu\Tz}\Bigr]^{\Tz=+\infty}_{\Tz=-\infty} . 
} 
Since the baryon number $N_B$ is calculated as 
$N_B = \int d^3x\, \langle J_B^0 \rangle = 1$, 
the baryon number density $\rho_B$ with respect to the radial direction 
$r = |\vecx -\vecX|$ is described as 
\eqn\BNdens{ 
\rho_B(r) = 4\pi r^2 \langle J_B^0 \rangle 
    = -4\pi r^2 \sum_{n=1}^\infty \biggl(\lambda_{2n-1}\Tkappa \int d\Tz\, 
    \Th(\Tz) \psi_{2n-1}(\Tz)\biggr) \psi_{2n-1}(\TZ) Y_{2n-1}(r) . 
} 
Then the isoscalar mean square radius becomes  
\eqnn\defisosMR 
$$\eqalignno{ 
\langle r^2 \rangle_{I=0} &= \int_0^\infty dr\, r^2 \rho_B(r) \cr 
&= 6\Tkappa\sum_{n=1}^\infty {1 \over \lambda_{2n-1}}\int d\Tz\, \Th(\Tz) \psi_{2n-1}(\Tz) \bigl\langle\psi_{2n-1}(\TZ)\bigr\rangle . &\defisosMR 
}$$ 
Since the baryon is almost localized at $\TZ=\TZ_{\rm cr} = 0$ on account 
of \barcri\ and \WFrhoZ{b}, $\bigl\langle\psi_{2n-1}(\TZ)\bigr\rangle$ 
can be approximated by $\psi_{2n-1}(0)$. Then, in the same way of \HSS, 
the isoscalar mean square radius is evaluated 
\eqn\isosMR{ 
\langle r^2 \rangle_{I=0} \approx {1 \over M_{\rm KK}^2}\int_0^\infty d\Tz' {1 \over \Tk(\Tz';\zeta)} \int_0^{\Tz'} d\Tz'' 6\Th(\Tz'';\zeta) , 
} 
where we recovered the factor $M_{\rm KK}$ explicitly. 
One can numerically compute these integration 
and depict the results depending 
on $\zeta$ in fig.~5. 
\bigskip 
\vbox{\centerline{\epsfbox{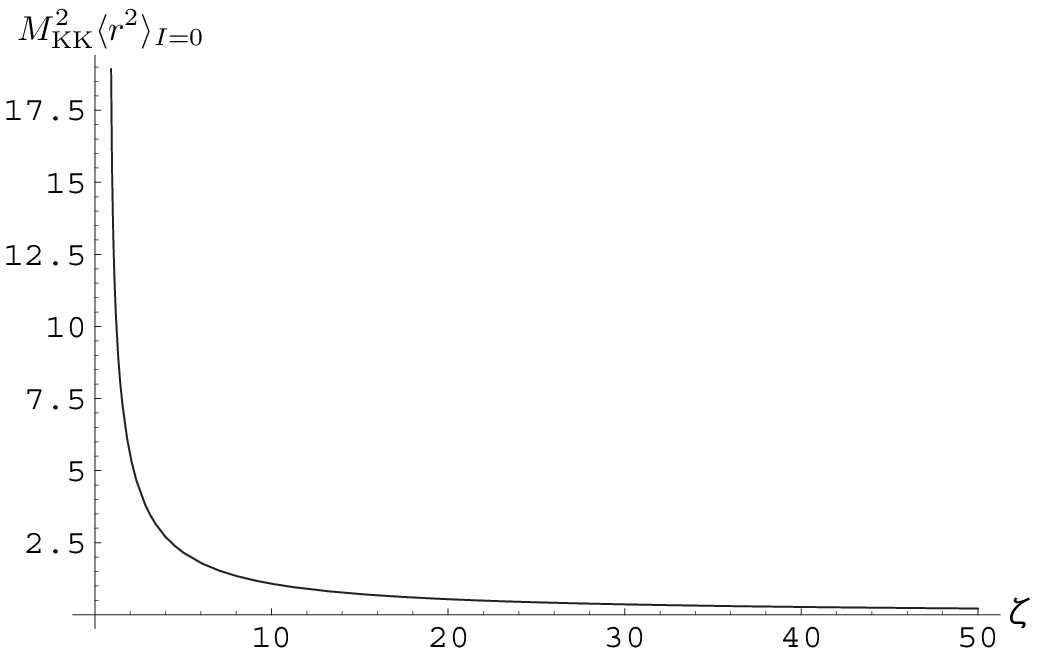}} 
\centerline{\fig\figisosMR{The $\zeta$ dependence of the isoscalar mean radius $M_{\rm KK}^2\langle r^2 \rangle_{I=0}$} The $\zeta$ dependence of the isoscalar mean radius $M_{\rm KK}^2\langle r^2 \rangle_{I=0}$}} 
\bigskip\nobreak\noindent 
The mean square radius \isosMR\ in the anti-podal case ($\zeta =1$) has 
been calculated in \HSS, that is, $M_{\rm KK}^2\langle r^2 
\rangle_{I=0} \approx 14.3$. 
If the mass scale $M_{\rm KK}$ is fixed, 
then the mean radius decreases with respect to $\zeta$ as can 
be seen in fig.~5. 
 
From the isovector charge $Q_V = (\tau_a/2)Q_V^a$, 
we obtain from \Vcur\ and \GFS{a} 
\eqnn\isovcha 
$$\eqalignno{ 
Q_V^a &= \tr \biggl(\tau^a \int d^3x\, J_V^0 \biggr) \cr 
    &= -\int dr\, 4\pi r^2 I^a 
    \sum_{n=1}^\infty \biggl(\lambda_{2n-1}\Tkappa \int d\Tz\, 
    \Th(\Tz) \psi_{2n-1}(\Tz)\biggr) \psi_{2n-1}(\TZ) Y_{2n-1}(r) , &\isovcha 
}$$ 
where we used 
$4\pi^2\Tkappa \rho^2 i \tr (\tau^a{\bf a}{\dot {\bf a}}^{-1})=I^a$, 
which is derived from \isosope. 
The isovector charge density $\rho_V(r)$ is defined 
by $Q_V^a = \int dr\, I^a \rho_V(r)$. 
Comparing \isovcha\ with \BNdens, 
we can show that $\rho_V(r)$ is equal to 
the baryon number density $\rho_B(r)$. 
So the isovector mean square charge radius is 
the same as the isoscalar mean square radius. 
This statement does not change from the $\zeta=1$ case 
investigated by \HSS. 
The electric mean square charge radii also have been mentioned in
\HSS, 
where the mean radius for a proton $\langle r^2\rangle_{E,p}$ 
and the one for a neutron $\langle r^2\rangle_{E,n}$ become 
$$ 
\langle r^2\rangle_{E,p} = \langle r^2\rangle_{I=0}, \quad 
\langle r^2\rangle_{E,n} = 0 . 
$$ 
These equations are satisfied also in the non-anti-podal case. 

Since the axial current \Acur\ leads to 
$$ 
\int d^3x J_A \propto -\int dr\, 4\pi r^2\sum_{n=1}^\infty \biggl(\lambda_{2n}\Tkappa\int d\Tz\, \Th(\Tz)\psi_{2n}(\Tz)\psi_0(\Tz)\biggr)\partial_\TZ \psi_{2n}(\TZ) Y_{2n}(r) 
={1 \over \Tk(\TZ)\xi(\infty)} , 
$$ 
and also $\int d^3x J_A \propto \int dr \rho_A(r)$, 
the axial charge density $\rho_A(r)$ is defined by 
$$ 
\rho_A(r) = {\Bigl\langle 4\pi r^2\sum_{n=1}^\infty \Bigl(\lambda_{2n}\Tkappa\int d\Tz\, \Th(\Tz)\psi_{2n}(\Tz)\psi_0(\Tz)\Bigr)\partial_\TZ \psi_{2n}(\TZ) Y_{2n}(r) \Bigr\rangle 
\over \Bigl\langle{1 \over \Tk(\TZ)\xi(\infty)}\Bigr\rangle } . 
$$ 
Now we shall approximate $\langle 1/\Tk(\TZ) \rangle$ by the 
classical value $1/\Tk(\TZ_{\rm cr}=0) = 1$. 
Then, in the way similar to \HSS, the axial radius 
$\langle r^2\rangle_A = \int dr\, r^2 \rho_A(r)$ is described as 
\eqn\axiMR{ 
\langle r^2\rangle_A = {3 \over M_{\rm KK}^2}\int_{-\infty}^\infty d\Tz {1 \over \Tk(\Tz)} 
    \int_0^\Tz d\Tz'\, \Th(\Tz')\psi_0(\Tz') . 
} 
In terms of \DefThk\ and \defxi\ we can numerically evaluate
$\langle r^2\rangle_A$, and its behavior is depicted in fig.~6.\foot{
The integrations in \axiMR\ are numerically done 
by Mathematica in terms of Monte-Carlo method.
} 
\bigskip 
\vbox{\centerline{\epsfbox{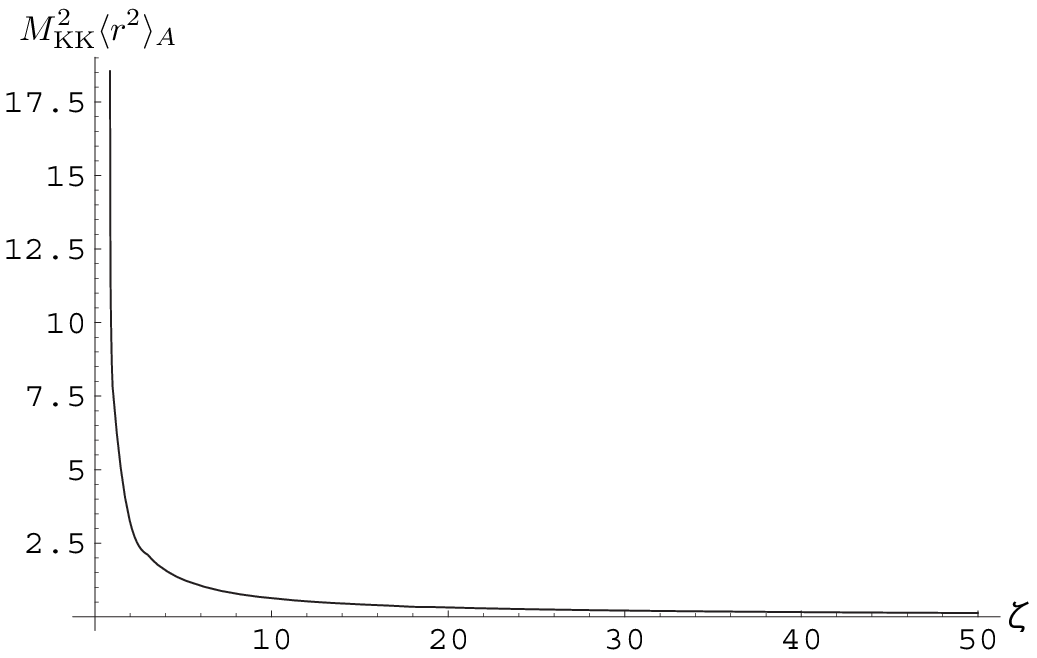}} 
\centerline{\fig\figaxiMR{} The $\zeta$ dependence of the axial charge mean redius $M_{\rm KK}^2\langle r^2\rangle_A$.}} 
\bigskip\nobreak\noindent 
From \figaxiMR, the axial charge mean radius is 
a monotonically decreasing function along $\zeta$. 
Ref.\HSS\ has calculated $M_{\rm KK}^2\langle r^2\rangle_A \approx
7.82$ 
in $\zeta=1$.

\medskip 
\noindent$\underline{\hbox{\it Magnetic moments}}$ 
\medskip 
In terms of the baryon number current \BNcur, 
the isoscalar magnetic moment is denoted by 
\eqn\defisosMM{ 
\mu_{I=0}^i = {1 \over 2}\epsilon^i{}_{jk}\int d^3x\, x^j J_B^k 
= -{\rho^2 \chi^i \over 4} , 
} 
where $\chi^i$ can be described from \Phidecomp\ and \spinope\ as 
$$ 
\chi^i = {1 \over 8\pi^2\Tkappa}J^i . 
$$ 
Here we concentrate on the up-spin proton and neutron states, 
which have the spin $(J^1,J^2,J^3) = (0,0,1/2)$ 
and the mass $M_N^{\rm exp} \approx 940$ [MeV]. 
By defining the $g$ factor as $\mu_{I=0}^i = g_{I=0} (\tau^i/4 M_N)$, 
we can identify the $g$ factor as 
\eqn\gfac{ 
g_{I=0} = {M_N \over 8\pi^2\Tkappa M_{\rm KK}} . 
} 
We should note that the $\zeta$-dependence is included in $\Tkappa$, 
which is determined through the pion decay constant 
$f_\pi^{\rm exp} \approx 92.4$ [MeV], 
\eqn\pidecconst{ 
\biggl({f_\pi^{\rm exp} \over 
M_{\rm KK}}\biggr)^2 = {4\Tkappa \over \pi^2}\int d\Tz {1 \over 
\Tk(\Tz;\zeta)} . 
} 
This equation is read from the mode expansion of 
\YMorig\ for the pion field \SS. 
The isoscalar magnetic moment $g_{I=0}$ can be rewritten as 
\eqn\isosgfac{ 
g_{I=0} = {M_{\rm KK}M_N \over 2\pi^4 f_\pi^2}\int_{-\infty}^\infty d\Tz {1 \over 
\Tk(\Tz;\zeta)} . 
} 
The $\zeta$-dependence of $g_{I=0}$ is 
proportional to $\int d\Tz\, \Tk(\Tz;\zeta)^{-1}$, 
which is depicted in fig.~7. 
\bigskip 
\vbox{\centerline{\epsfbox{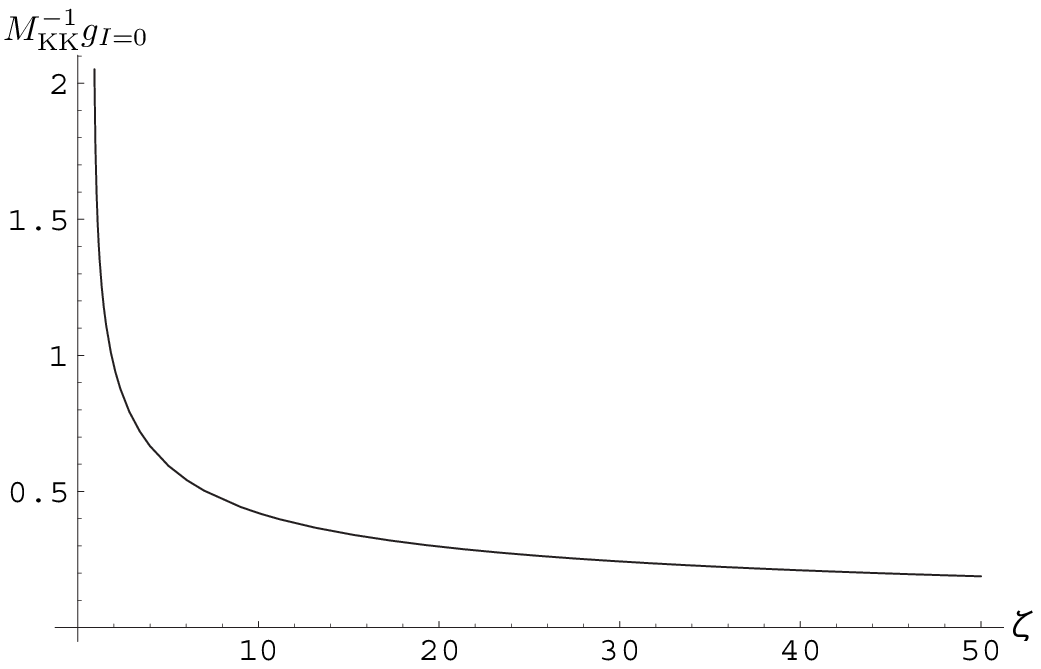}} 
\centerline{\fig\figisosMo{} The plot of $M_{\rm KK}^{-1} g_{I=0}(\zeta)$.}} 
\bigskip\nobreak\noindent 
 
The isovector magnetic moment is given by 
\eqn\defisovMM{ 
\mu_{I=1}^i = \epsilon^i{}_{jk} \int d^3x\, x^j\tr(J_V^k \tau^3) 
    = -4\pi^2\Tkappa\rho^2 \tr ({\bf a}\tau^i{\bf a}^{-1}\tau^3) , 
} 
We can evaluate \defisovMM\ for the up-spin proton and neutron states as 
$$ 
\langle \mu_{I=1}^i \rangle_{p} = -\langle \mu_{I=1}^i \rangle_{n} 
= {8\pi^2\Tkappa \over 3}\langle \rho^2 \rangle \delta^{3i} . 
$$ 
$\langle \rho^2 \rangle$ is calculated 
in terms of the wave function \WFrhoZ{a} 
$$ 
\langle \rho^2 \rangle = {\int d\rho\, \rho^5 R(\rho)^2 \over \int d\rho\, \rho^3 R(\rho)^2} 
= {\sqrt{5} +2\sqrt{5+N_c^2} \over 2N_c}\rho_{\rm cr}^2(\zeta) , 
$$ 
where $\rho_{\rm cr}$ has been calculated in \barcri. 
Since the $g_{I=1}$ factor is defined in the same way 
as the isovector magnetic moment, we obtain 
\eqn\isovgfac{ 
g_{I=1} = {2\sqrt{2} M_N \over M_{\rm KK}}\biggl(1+2\sqrt{1+{N_c^2 \over 5}}\biggr){\zeta \over \sqrt{8\zeta^3 -5}} . 
} 
The function $M_{\rm KK}g_{I=1}$ of $\zeta$ with $N_c = 3$ is 
drawn in the following figure: 
\bigskip 
\vbox{\centerline{\epsfbox{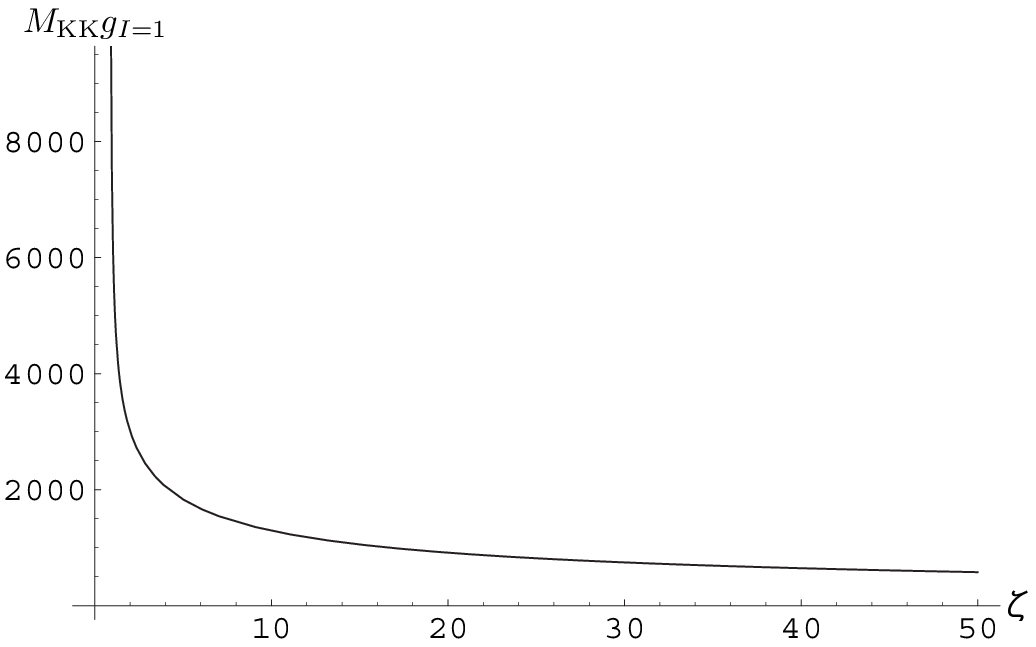}} 
\centerline{\fig\figisovMo{} The plot of $M_{\rm KK}g_{I=1}(\zeta)$ with $M_N=940$ and $N_c = 3$.}} 
\bigskip\nobreak\noindent 
By the use of \isosgfac\ and \isovgfac, the magnetic moments for a proton 
and a neutron can be easily computed as $\mu_p = (g_{I=0}+g_{I=1})/4$ and 
$\mu_n = (g_{I=0}-g_{I=1})/4$ respectively.

\medskip 
\noindent$\underline{\hbox{\it Couplings}}$ 
\medskip 

The axial coupling $g_A$ is defined 
in terms of the axial current $J_A^i$ in \Acur\ as 
\eqn\defaxiCu{ 
\int d^3x \langle J_A^{a,i}\rangle = {1 \over 2}g_A \langle \tr({\bf a}\tau^i{\bf a}^{-1}\tau^a) \rangle , 
} 
where $J_A^{a,i} = \tr (\tau^a J_A^i)$. 
Since the left hand side of \defaxiCu\ is calculated 
from \LRcur, \Acur\ and \GFS{b}, 
$$ 
\int d^3x \langle J_A^{a,i}\rangle = {8\pi^2 \Tkappa \over 6\xi(\infty)}\biggl\langle{\rho^2 \over \Tk(\TZ)}\biggr\rangle \langle \tr({\bf a}\tau^i{\bf a}^{-1}\tau^a) \rangle , 
$$ 
one can read the axial coupling 
\eqn\axiCu{ 
g_A(\zeta) = {8\pi^2 \Tkappa \over 3 \xi(\infty)}\biggl\langle{\rho^2 \over \Tk(\TZ)}\biggr\rangle 
\approx {\sqrt{2}N_c \over \xi(\infty)}{\zeta \over \sqrt{40\zeta^3 -25}} . 
} 
The approximation was given by the classical values, that is, 
$\rho \approx \rho_{\rm cr}$ and $\Tk(\TZ) \approx \Tk(\TZ_{\rm cr})=1$ 
with \barcri. 
We should note that $\xi(\infty)$ also depends on $\zeta$ 
and can be numerically computed from \defxi. 
Then $g_A(\zeta)$ with $N_c = 3$ can be drawn as fig.~9. 
\bigskip 
\vbox{\centerline{\epsfbox{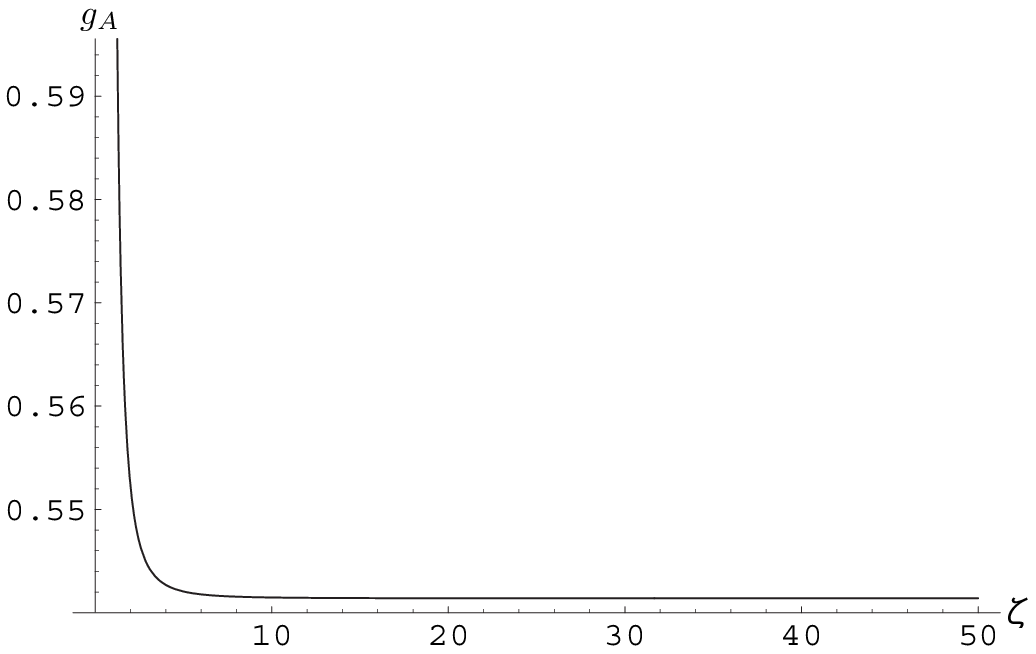}} 
\centerline{\fig\figaxiCu{} The plot of $g_A(\zeta)$ with $N_c = 3$.}} 
\bigskip\nobreak\noindent 
In the anti-podal case, \HSS\ evalulated the axial coupling, 
$g_A(\zeta=1)\approx 0.697$. 
Since \axiCu\ is independent of $M_{\rm KK}$, 
we shall compare $g_A(\zeta)$ with the experimental datum $g_A^{\rm
exp}$, 
which is approximately equal to $1.27$. 
If we set $g_A(\zeta) \approx 1.27$, 
then \axiCu\ leads to $\zeta\approx 0.870$. 
But this solution is nonsense, 
because $\zeta$ must be in $[1,\infty)$ by definition. 
At present, the best fitted value of $\zeta$ for the experimental data 
is $\zeta=1$, that is, the anti-podal SS model. 
We shall give more comments on this issue in Section 7.

%%%%%%%%%%%%%%%%%%%%%%%%%%%%%% 
\subsec{$M_{\rm KK}$ and $\zeta$ fitted to experimental data} 
 
So far we have calculated the baryon mass spectra, 
the mean radii, the magnetic moments and the couplings 
as functions of $M_{\rm KK}$ and $\zeta$. 
Comparing those quantities with the experimental data, 
we shall determine $M_{\rm KK}$ and $\zeta$. 
We remind the reader that for $\zeta=1$ 
these properties of the baryons were computed in \HSS. 
The idea is to find the values of $M_{\rm KK}$ and $\zeta$ 
that yield the best fit to the experimental data. 
We shall extract the relation between $M_{\rm KK}$ and $\zeta$ 
in five different ways from the baryonic data 
and in two more ways from the mesonic spectra. 

We start with the mass difference \estI\ 
between the neucleon, $n(940)$, and 
the lowest mode of $\Delta$-baryon, $\Delta(1232)$, 
from which we find the following relation 
between $M_{\rm KK}$ and $\zeta$: 
\eqn\Mkkbar{ 
M_{\rm KK} = {876 \sqrt{5} \over 
\sqrt{58}-\sqrt{28}}{\zeta \over \sqrt{8\zeta^3-5}} =: {\cal 
B}_1(\zeta) . 
} 
 
Next we use  the isoscalar mean square radius \isosMR\ to obtain 
\eqn\MkkisosMR{ 
M_{\rm KK} = \sqrt{{6 \over \langle r^2 
\rangle_{I=0}^{\rm exp} }\int_0^\infty d\Tz' \Tk(\Tz';\zeta)^{-1} 
\int_0^{\Tz'} d\Tz'' \Th(\Tz'';\zeta)} =: {\cal B}_2(\zeta) , 
} 
where the experimental datum of the isoscalar mean square radius 
$\langle r^2 \rangle_{I=0}^{\rm exp} \approx 0.806$ [fm]. 
 
Using  \defxi, we calculate $M_{\rm KK}$ from the axial mean radius \axiMR, 
\eqn\MkkaxiMR{ 
M_{\rm KK}  = \sqrt{{ 3 \int_{-\infty}^\infty 
d\Tz\, \Tk(\Tz)^{-1} 
    \int_0^\Tz d\Tz'\, \Th(\Tz') \int_0^{\Tz'} d\Tz''\, \Tk(\Tz'')^{-1} \over 
\langle r^2\rangle_A^{\rm exp} \int_0^\infty d\Tz\, \Tk(\Tz)^{-1}} } 
=: {\cal B}_3(\zeta) , 
} 
where the experimental datum of the axial mean square radius 
$\langle r^2\rangle_A^{\rm exp} \approx 0.674$ [fm]. 
 
The isoscalar magnetic moment \isosgfac\ yields the relation 
\eqn\Mkkiscag{ 
M_{\rm KK} = {\pi^4 (f_\pi^{\rm exp})^2 g_{I=0}^{\rm 
exp} \over 
    M_N^{\rm exp} \int_0^\infty d\Tz\, \Tk(\Tz;\zeta)^{-1} } 
=: {\cal B}_4(\zeta) . 
} 
The experimental data of the pion decay constant and 
the isoscalar magnetic moment are given by 
$f_\pi^{\rm exp} \approx 92.4$ [MeV] and 
$g_{I=0}^{\rm exp} \approx 1.76$. 
Substituting $N_c=3$ into the isovector magnetic moment \isovgfac, 
$M_{\rm KK}$ is written down as 
\eqn\Mkkivecg{ 
M_{\rm KK} = {2\sqrt{2} M_N^{\rm exp} \over g_{I=1}^{\rm exp}}\biggl(1+2\sqrt{{14 \over 5}}\biggr){\zeta \over \sqrt{8\zeta^3 -5}} 
=: {\cal B}_5(\zeta) . 
} 
$M_N^{\rm exp}$ and $g_{I=1}^{\rm exp}$ are given 
by the experimental values, $M_N^{\rm exp} \approx 940$ [MeV] 
and $g_{I=1}^{\rm exp} \approx 9.41$. 
 
The meson spectra have been studied extensively in the literature. 
Here we consider the $\rho$ and $a_1$ mesons. 
We match the calculated masses with the experimental data, so that 
\eqnn\Mkkrho 
\eqnn\Mkkaone 
$$\eqalignno{ 
M_{\rm KK} &= {m_\rho^{\rm exp} \over m_\rho(\zeta)} =: {\cal 
M}_1(\zeta) , &\Mkkrho \cr 
M_{\rm KK} &= {m_{a_1}^{\rm exp} \over 
m_{a_1}(\zeta)} =: {\cal M}_2(\zeta) . &\Mkkaone 
}$$ 
The functions $m_\rho(\zeta)$ and $m_{a_1}(\zeta)$ are dimensionless 
and have been evaluated numerically in \PZ. 
The experimental values of the meson masses are described 
as $m_\rho^{\rm exp} \approx 776$ [MeV] and 
$m_{a_1}^{\rm exp} \approx 1230$ [MeV]. 
\bigskip 
\vbox{\centerline{\epsfbox{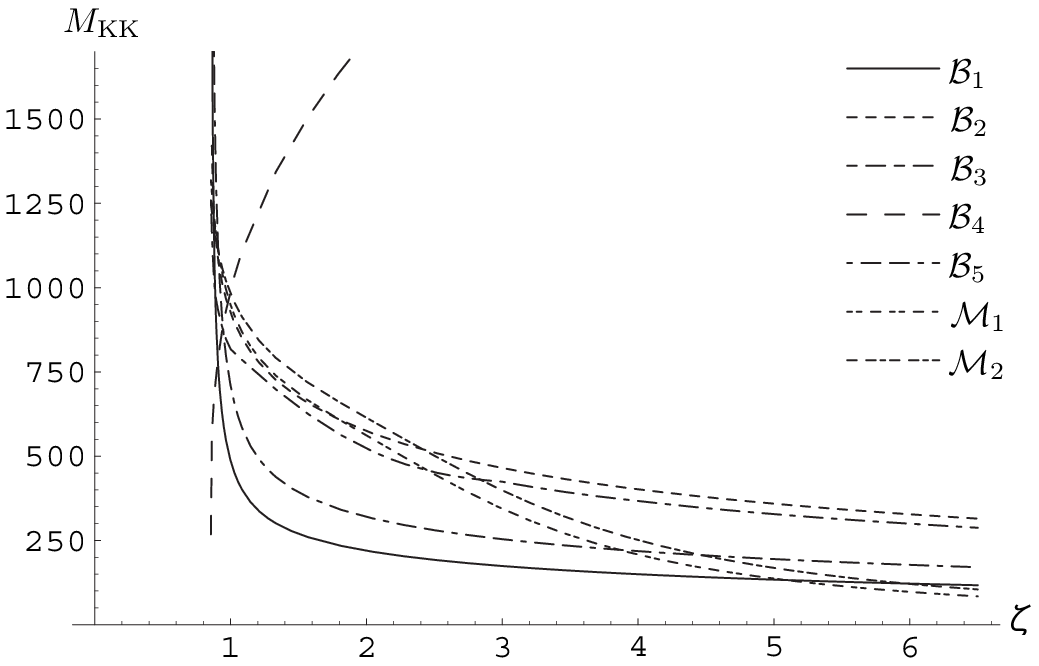} 
} 
\centerline{\fig\figmkkzeta{} The behaviors of $M_{\rm KK}$ on $\zeta$.}} 
\bigskip\nobreak\noindent 
\vbox{\centerline{\epsfbox{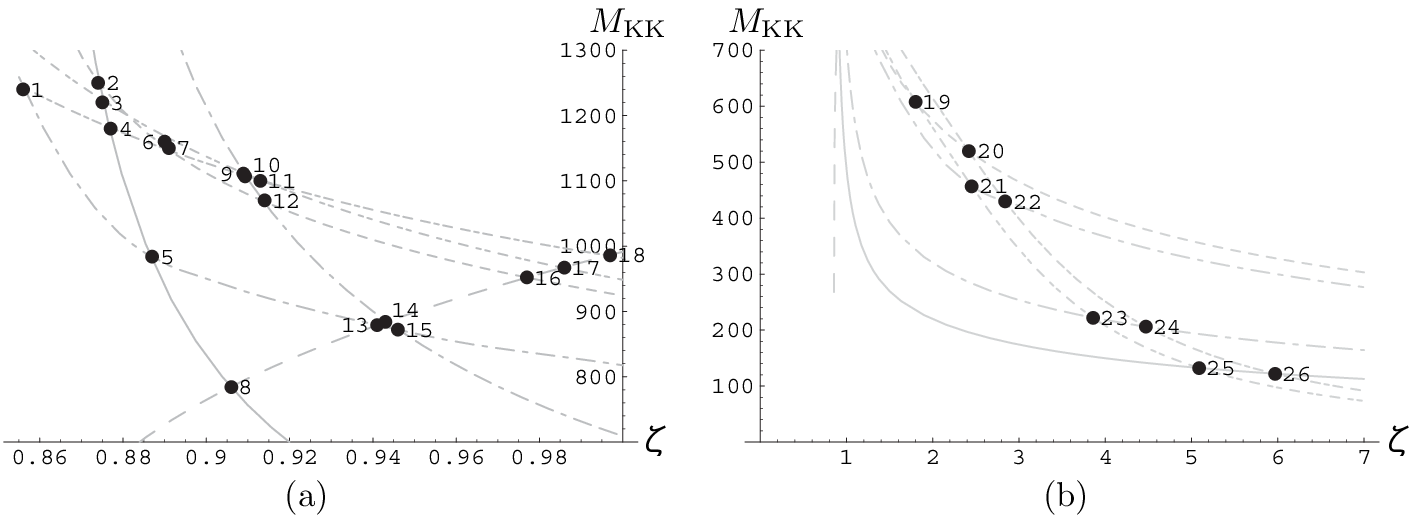}} 
\centerline{\fig\figcrosspt{} (a) The crossing points in $\zeta<1$. (b) The crossing points in $\zeta \geq 1$.}} 
\bigskip

The various forms of dependence of $M_{\rm KK}$ on $\zeta$ are 
depicted in \figmkkzeta. 
In \figcrosspt\ we enlarge the picture in the two regions 
where the various functions are crossing. 
One region (\figcrosspt\ (a)) is in the ``un-physical domain'' 
where $\zeta<1$, and the other is for $\zeta\geq 1$. 
The values $(M_{\rm KK},\zeta)$ of each crossing point in \figcrosspt\ 
is listed in the following table: 
\bigskip
\centerline{\vbox{\offinterlineskip 
\halign{ 
\hskip2pt \hfil#\strut\hfil \hskip1pc 
& #\strut\hfil 
& \vrule# \hskip3pt 
& \hfil#\strut\hfil \hskip1pc 
& #\strut\hfil 
&\hskip2pt \vrule#\hskip1pt\vrule \hskip3pt 
& \hfil#\strut\hfil \hskip1pc 
& #\strut\hfil \hskip1pt \cr 
\noalign{\hrule} 
label & $(\zeta, M_{\rm KK})$ && label & $(\zeta, M_{\rm KK})$ && label & $(\zeta, M_{\rm KK})$ \cr 
\noalign{\hrule} 
1 & $(0.856, 1240)$ && 10 & $(0.909, 1110)$ && 
19 & $(1.80, 608)$ \cr 
2 & $(0.874, 1250)$ && 11 & $(0.913, 1100)$ && 
20 & $(2.42, 520)$ \cr 
3 & $(0.875, 1220)$ && 12 & $(0.914, 1070)$ && 
21 & $(2.45, 457)$ \cr 
4 & $(0.877, 1180)$ && 13 & $(0.941, 879)$ && 
22 & $(2.84, 430)$ \cr 
5 & $(0.887, 984)$ && 14 & $(0.943, 884)$ && 
23 & $(3.86, 222)$ \cr 
6 & $(0.890, 1160)$ && 15 & $(0.946, 872)$ && 
24 & $(4.47, 206)$ \cr 
7 & $(0.891, 1150)$ && 16 & $(0.977, 952)$ && 
25 & $(5.09, 132)$ \cr 
8 & $(0.906, 784)$ && 17 & $(0.986, 967)$ && 
26 & $(5.97, 122)$ \cr 
9 & $(0.909, 1110)$ && 18 & $(0.997, 986)$ && 
 &  \cr 
\noalign{\hrule} 
} 
}}\nobreak 
\centerline{Table 4: The crossing points in \figcrosspt.} 
\bigskip\noindent 

\Mkkbar\ and \Mkkivecg\ have no crossing point 
and ${\cal B}_5/{\cal B}_1$ is independent of $\zeta$. 
${\cal B}_5/{\cal B}_1$ should be equal to one in order 
for the prediction of the model to fit the obsevational values. 
In fact, substituting the experimental values, we evaluate 
$$ 
{\cal B}_5/{\cal B}_1 = {(\sqrt{29}-\sqrt{14})(\sqrt{5}+2\sqrt{14}) g_{I=1}^{\rm exp} \over 1095 M_N^{\rm exp}} 
\approx 1.457 , 
$$ 
which means 45.7\% difference. 
 
A better way to determine the values of the two parameters  
$(\zeta, M_{\rm KK})$ is by a fit of the calculated results 
to the experimental data using a  $\chi^2$-method. 
This leads to the values 
\eqn\bstmkkzeta{ 
\zeta = 0.942 ,\quad M_{\rm KK} = 997\ {\rm [MeV]} . 
} 
 
Since in the model of \SS\ $\zeta$ must satisfy 
$\zeta \geq 1$ by definition, 
the result for $\zeta$ in \bstmkkzeta\ does not make sense. 
Let us ignore this problem for a moment, estimate the 
physical quantities naively by using the values \bstmkkzeta\ and 
then discuss possible scenario that yields this situation. 
The calculated results based on  \bstmkkzeta\ are summarized in Table 5. 
\bigskip 
\centerline{\vbox{\offinterlineskip 
\halign{ 
\hskip1pt \hfil#\strut\hfil \hskip1pc 
& \hfil#\strut\hfil \hskip1pc 
& \hfil#\strut\hfil \hskip1pc 
& \hfil#\strut\hfil \hskip1pt \cr 
\noalign{\hrule} 
 & our model & experiment & discrepancy[\%]  \cr 
\noalign{\hrule \vskip1pt \hrule} 
$m_\rho$ & 746 MeV & 776 MeV & $-3.86$ \cr 
$m_{a_1}$ & 1160 MeV & 1230 MeV & $-5.31$ \cr 
${m_{\Delta(1232)} \over m_{n(940)}}$ & 1.51 & 1.31 & $15.2$ \cr 
$\sqrt{\langle r^2 \rangle_{I=0}}$ & 0.813 fm & 0.806 fm & $0.920$ \cr 
$\sqrt{\langle r^2 \rangle_A}$ & 0.594 fm & 0.674 fm & $-11.9$ \cr 
$g_{I=0}$ & 1.99 & 1.76 & $13.1$ \cr 
$g_{I=1}$ & 8.41 & 9.41 & $-10.7$ \cr 
\noalign{\hrule} 
} 
}}\nobreak 
\centerline{Table 5: The $\chi^2$-fitting. } 
\bigskip\noindent 
Note that, in Table 5, we fixed $m_{n(940)}=940$ 
in the calculation of $m_{\Delta(1232)} / m_{n(940)}$. 
The axial coupling $g_A$ is independent of $M_{\rm KK}$, 
and is evaluated in terms of $\zeta$ in \bstmkkzeta\ as $g_A = 0.779$, 
which has $-38.7$\% difference from the experimental value. 

Now let us come back to the issue of possible meaning of \bstmkkzeta. 
First notice that the value of $\zeta$ is larger 
than the critical value defined in Section 2, $\zeta_{cr}=(5/8)^{1/3}$. 
The fact that the value of $\zeta$ yielding the best fit came out 
to be in the un-physical region of $\zeta<1$ 
may indicate that the description of the baryonic phenomena 
in the model \SS, as given in \HSS, has to be modified. 
We cannot pinpoint the precise reason for that, 
but it might be that, due to local back reaction 
of the flavor brane with the baryon vertex on the background, 
the U-shaped cigar geometry is distorted 
so that effectively $\zeta<1$ is allowed.
Again we do not know that this is indeed the case 
but it seems to us that the fact that 
we have found the parameter $\zeta$ out of its region of definition 
may indicate a problem with the scenario for \bstmkkzeta.

%%%%%%%%%%%%%%%%%%%%%%%%%%%%%%  
\newsec{Baryons in single flavor model $(N_f = 1)$} 

We have started our journey with the baryon vertex 
attached to the flavor branes with $N_c$ strings. 
In this picture, which was analyzed in Section 3, 
nothing forbids us from taking only one single flavor brane, namely $N_f=1$. 
The heuristic arguments about the stability of the configuration 
apply also to the single flavor brane case, 
and moreover the conclusion that the baryon vertex is 
immersed in the flavor brane and does not hang out of it 
applies here as well. 
Thus, we conclude that there should be baryonic solutions 
for the abelian analog of \YMact\ plus \CSact. 
In fact from the point of view of the underlying $SU(N_c)$ QCD theory, 
there is no reason that there will not exist baryonic states 
as singlets of the gauge symmetry composed from $N_c$ quarks. 
 
The action describing the theory on the single flavor, which is  
reduced from the expanded DBI action and the Chern-Simon term, 
takes the following form: 
\eqn\oneFact{ 
S_{N_f=1} = -\Tkappa \int d^4x d\Tz 
\biggl({1 \over 2}\Th(\Tz)F_{\mu\nu}^2 +\Tk(\Tz)F_{\mu \Tz}^2\biggr) 
    +{9\pi \kappa \over 4 \lambda}\int d^4x d\Tz\, \epsilon^{ijk}A_0F_{ij}F_{k\Tz} . 
} 
Here $(A_\mu,A_\Tz)$ denotes a $U(1)$ gauge field in five dimensions.

The associated equations of motion are 
\eqna\EOMoneF 
$$\eqalignno{ 
\Th(\Tz)\partial_iF^{i0} +\partial_\Tz\bigl(\Tk(\Tz)F^{\Tz 0}\bigr) 
    &= -{9\pi\kappa \over 8\lambda\Tkappa}\epsilon^{ijk}F_{ij}F_{k\Tz} , &\EOMoneF{a} \cr 
\Th(\Tz)\partial_\mu F^{\mu i} +\partial_\Tz\bigl(\Tk(\Tz)F^{\Tz 
i}\bigr) 
    &= -{9\pi\kappa \over 8\lambda\Tkappa}\epsilon^{ijk}[2\partial_j(A_0F_{k\Tz}) +\partial_\Tz(A_0F_{jk})] , &\EOMoneF{b} \cr 
\Tk(\Tz)\partial_\mu F^{\mu \Tz} 
    &= {9\pi\kappa \over 8\lambda\Tkappa}\epsilon^{ijk}\partial_k(A_0F_{ij}) . &\EOMoneF{c} 
}$$ 
For simplicity, we shall consider the anti-podal case $(\zeta =1)$, 
in which $\Tz =z$, $\Th(\Tz) = (1+z^2)^{-1/3}$, 
$\Tk(\Tz) = 1+z^2$ and $\Tkappa = \kappa$. 
We assume that the $U(1)$ gauge field is static 
and analyze the leading behavior in the $\lambda^{-1}$ expansion 
under the rescaling \lamscale. 
Then the equations of motion \EOMoneF{} are reduced to 
\eqna\lamEoneF 
$$\eqalignno{ 
\partial_M^2 A_0 
    &= -{9\pi \over 8}\epsilon^{ijk}F_{ij}F_{kz} , &\lamEoneF{a} \cr 
\partial_M F^{MN} &= 0 . &\lamEoneF{b} 
}$$ 
\lamEoneF{b} is the $U(1)$ version of the instanton equation 
in four-dimensional Euclidean space. 
Now it is well known that the abelian theory does not admit 
a non-singular instanton solution and 
thus we are facing a problem of how to identify the baryon 
in such a theory. 
In fact this situation is of no surprise, 
since in a similar manner there is no Skyrmion solution 
to an abelian Skyrme-like theory. 

We suspect that there should be a solution 
once we switch back the curvature nature of the five-dimensional model, 
namely when we include higher order corrections in $1/\lambda$. 
This is an open question that deserves a further study.

%%%%%%%%%%%%%%%%%%%%%%%%%%%%%% 
\newsec{Baryons in six-dimensional holographic model} 

In analogy to SS model \SS, one can introduce 
a stack of $N_f$ D4-branes and a stack of $N_f$ anti-D4-branes 
to the background of near extremal D4-branes 
of a six-dimensional non-critical gravity model \refs{\KSii,\KSi}. 
The model, which like all other non-critical models 
suffers from the fact that it has order one curvature, 
is based on a compactified $AdS_6$ spacetime 
with a constant dilaton 
and hence does not suffer from large string coupling 
as happens in SS model. 
The spectra of mesons were analyzed in \refs{\CPS,\MB} 
and its thermal phase structure was determined in \MS. 
Most of the properties of the non-critical holographic model 
are similar to those of SS model, 
but some properties like the dependence of the meson masses 
on the stringy mass of the quarks 
and the excitation number are different. 
 
The purpose of this section is 
to investigate the baryon configurations 
in the non-critical holographic model of \KSi\ and 
to see how, if at all, it differs from those of the critical model. 
As was discussed in Section 3, 
the baryon vertex is a D4-brane wrapping the transverse $S^4$ cycle. 
In the six dimensional model by construction the $S^4$ 
does not exist, so one may wonder that 
the whole idea might not work for that model. 
However, one can use instead unwrapped D0-branes. 
In analogy to the Chern-Simon term on the worldvolume 
of the wrapped D4-branes discussed in Section 3, 
there is also a Chern-Simon term of the form $N_c A_0 dt$ 
on the D0-brane worldvolume 
and hence also in this case one needs to attach $N_c$ strings 
to the D0-baryon vertex. 
The other end of each of these strings will be obviously attached 
to the probe flavor D4-branes. 
Just as for the near extremal D4-branes of the critical model, 
and in fact as is shown in appendix A for any D$p$ branes, 
also in the non-critical model the baryon vertex will be attached 
to the probe branes. 
Let us now analyze the baryons in the corresponding 
five-dimensional theory. 
 
The background of this model \refs{\KSii,\KSi} is given by 
\eqnn\NCmet 
$$\eqalignno{ 
&ds^2 = \biggl({u \over R}\biggr)^{2}\bigl[\eta_{\mu\nu}dx^\mu 
dx^\nu +f(u) dx_4^2 \bigr] +\biggl({R \over u}\biggr)^{2}{du^2 \over 
f(u)} , &\NCmet \cr &e^\phi = {2\sqrt{2} \over \sqrt{3}N_c} ,\quad 
F_{(6)} = -N_c \biggl({u \over R}\biggr)^{4} dx_0 \wedge dx_1 \wedge 
dx_2 \wedge dx_3 \wedge dx_4 \wedge du , \cr &R^2 = {15 \over 2} 
,\quad f(u) := 1 - \biggl({u_{\rm KK} \over u}\biggr)^5 . 
}$$ 
Since the period of $x_4$ direction is $4\pi R^2 / (5u_{\rm KK})$, 
the mass scale is 
$$ 
M_{\rm KK} = {5 u_{\rm KK} \over 2R^2} . 
$$ 
We concentrate on the $N_f =2$ case and use the same decomposition 
of the $U(2)$ gauge field as in \decgauge. 
In this background \NCmet, the action of the flavor D4-branes 
is described by 
$$\eqalign{ 
S &= T_4 \int d^5x\, e^{-\phi} \sqrt{-\det(g_{MN} +2\pi\alpha' 
\CF_{MN})} +T_4 {\tilde a} \int {\cal P}(C_{(5)}) + b \int 
\omega_5^{U(2)} \cr &= S_0 + S_{\rm YM} + S_{\rm CS} + \CO(\CA^3) , 
}$$ 
where 
\eqnn\NCYM 
\eqnn\NCCS 
$$\eqalignno{ 
S_0 &= T_4 e^{-\phi} \int d^4x dx_4 \biggl({u \over R}\biggr)^5 
\Biggl[ 
    \sqrt{f(u) + \biggl({R \over u}\biggr)^4{u'^2 \over f(u)}} - a \Biggr] , \cr 
S_{\rm YM} &= -{\tilde T} \int d^4x dz\, 
    \tr\biggl[{1 \over 2}h(z)\eta^{\mu\nu}\eta^{\rho\sigma}F_{\mu\rho}F_{\nu\sigma} +M_{\rm KK}^2 k(z)\eta^{\mu\nu}F_{\mu z}F_{\nu z} \biggr] , &\NCYM\cr 
%& {\tilde T} := {(2\pi\alpha')^2 T_4 R \over 4e^{\phi} u_{\rm KK}} =: c N_c , \cr 
S_{\rm CS} &= b\epsilon^{MNPQ} \int d^4x dz \biggl[ {3 \over 8}\HA_0 
\tr(F_{MN}F_{PQ}) -{3 \over 2}\HA_M \tr(\partial_0A_N F_{PQ}) \cr 
&\quad  +{3 \over 4}\HF_{MN} \tr(A_0F_{PQ}) +{1 \over 
16}\HA_0\HF_{MN}\HF_{PQ} -{1 \over 4}\HA_M\HF_{0N}\HF_{PQ} \biggr] , 
&\NCCS 
}$$ 
up to total derivatives. 
${\tilde T}$ is equal to 
${(\pi\alpha')^2 T_4 R e^{-\phi} {u_{\rm KK}}^{-1}}$, 
which is proportional to $N_c$. 
So we describe ${\tilde T} := c N_c$. 
Note that ${\tilde a},b$ are constants 
and $a=(2/\sqrt{5}){\tilde a}$ \MS. 
Introducing the coordinate $z$ defined by 
$$ 
\biggl({u \over u_{\rm KK}}\biggr)^5 = \zeta^5 + \zeta^3 z^2 , \quad 
\zeta := {u_0 \over u_{\rm KK}} , 
$$ 
we compute $h(z)$ and $k(z)$ in the power expansion for small $z$, 
$$\eqalign{ 
&h(z) =  h_0 +h_1 z^2 +\CO(z^4) , \quad k(z) = k_0 +k_1 z^2 
+\CO(z^4) , \cr 
&h_0 = {4 \zeta^{3 \over 2}\over 5 \sqrt{2\zeta^5 -1 
-2a\zeta^{5/2}\sqrt{\zeta^{5} -1}}} ,\quad h_1 = {2(a^2-1) \zeta^{9 
\over 2} \over 5 \Bigl(2\zeta^5 -1 -2a\zeta^{5/2}\sqrt{\zeta^5 
-1}\Bigr)^{3/2}} ,\cr 
&k_0 = {4 \over 5}\zeta^{1 \over 2} 
\sqrt{2\zeta^5 -1 -2a \zeta^{5/2}\sqrt{\zeta^5 -1}} ,\quad k_1 = {2 
\over 25}{(13-5a^2)\zeta^5 -4 -8a\zeta^{5/2}\sqrt{\zeta^5 -1} \over 
\zeta^{3/2}\sqrt{2\zeta^5 -1 -2a\zeta^{5/2}\sqrt{\zeta^5 -1}}} . 
}$$ 
Without any loss of generality, we can set $M_{\rm KK} = 1$ again. 
Using the rescaling  
\eqn\ncscale{\eqalign{ 
&x^0 \to x^0,\quad x^i 
\to {1 \over \sqrt{N_c}}x^i,\quad z \to {1 \over \sqrt{N_c}}z , \cr 
&\CA_0 \to \CA_0,\quad \CA_i \to \sqrt{N_c}\CA_i,\quad \CA_z \to 
\sqrt{N_c}\CA_z , 
}} 
the Yang-Mills action \NCYM\ is expanded with respect to the large $N_c$, 
$$\eqalign{ 
S_{\rm YM} &= -c \int d^4x dz\, \tr \biggl[ 
    N_c\biggl({1 \over 2}h_0 F_{ij}^2 +k_0 F_{iz}^2\biggr) \cr 
&\qquad +{1 \over 2}h_1 z^2 F_{ij}^2 +k_1 z^2 F_{iz}^2 -h_0 F_{0i}^2 
-k_0 F_{0z}^2 
    + \CO(N_c^{-1}) \biggr] \cr 
&\quad -c \int d^4x dz\, {1 \over 2} \biggl[ 
    N_c\biggl({1 \over 2}h_0 \HF_{ij}^2 +k_0 \HF_{iz}^2 \biggr) \cr 
&\qquad +{1 \over 2}h_1 z^2 \HF_{ij}^2 +k_1 z^2 \HF_{iz}^2 -h_0 
\HF_{0i}^2 -k_0 \HF_{0z}^2 
    + \CO(N_c^{-1}) \biggr] . 
}$$ 
Then the equations of motion for the $SU(2)$ part are described as 
\eqna\NCeomSU 
$$\eqalignno{ 
&h_0 D^i F_{i0} + k_0 D^z F_{z0} - {3b \over 
8c}\epsilon^{MNPQ}\HF_{MN}F_{PQ} = 0 , &\NCeomSU{a} \cr 
&h_0 D^i 
F_{ij} + k_0 D^z F_{zj} = 0 , &\NCeomSU{b} \cr 
&k_0 D^i F_{iz} = 0 , 
&\NCeomSU{c} 
}$$ 
while the equations of motion for the $U(1)$ part are 
\eqna\NCeomU 
$$\eqalignno{ 
&h_0 \partial^i \HF_{i0} + k_0 \partial^z \HF_{z0} - {3b \over 
8c}\epsilon^{MNPQ}\biggl[ \tr (F_{MN}F_{PQ}) +{1 \over 
2}\HF_{MN}\HF_{PQ} \biggr] = 0 , &\NCeomU{a} \cr 
&h_0 \partial^i \HF_{ij} + k_0 \partial^z \HF_{zj} = 0 , &\NCeomU{b} \cr 
&k_0 \partial^i \HF_{iz} = 0 . &\NCeomU{c} 
}$$ 
Since \NCeomSU{b,c} correspond to the instanton equation, 
in completely the same way as in SS model, 
the equations of motion \NCeomSU{} and \NCeomU{} 
can be solved as 
\eqna\NCsol 
$$\eqalignno{ 
&A_M(x^i,z) = -iv(\xi)g\partial_Mg^{-1} \quad (M=1,2,3,z), &\NCsol{a} \cr 
&A_0 = 0 , &\NCsol{b} \cr 
&\HA_M = 0, &\NCsol{c} \cr 
&\HA_0 = {3b \over c\sqrt{h_0k_0}}{1 \over \xi^2}\biggl[ 
    1-{\rho^4 \over (\xi^2 + \rho^2)^2}\biggr] , &\NCsol{d} 
}$$ 
where 
$$\eqalign{ 
&v(\xi) = {\xi^2 \over \xi^2 + \rho^2} ,\quad g(x^i,z) = {s(z-Z){\bf 
1} -i(x^i-X^i)\tau_i \over \xi} , \cr &\xi := \sqrt{(x^i - X^i)^2 + 
s^2 (z-Z)^2} ,\quad s := \sqrt{h_0 \over k_0} . \cr 
}$$ 
These solutions \NCsol{} lead to the baryon mass, 
$$\eqalign{ 
M &= N_c c \int d^3x dz \,\tr \biggl({h_0 \over 2} F_{ij}^2 +k_0 
F_{iz}^2\biggr) \cr &\quad  + c \int d^3x dz \biggl[\tr\biggl( {h_1 
\over 2} z^2 F_{ij}^2 +k_1 z^2 F_{iz}^2 \biggr) 
    -{h_0 \over 2}(\partial_i\HA_0)^2 -{k_0 \over 2}(\partial_z\HA_0)^2 \cr 
&\qquad -{3b \over 8c} \epsilon^{MNPQ}\HA_0\tr(F_{MN}F_{PQ}) \biggr] 
+\CO(N_c^{-1}) \cr 
%&= c\biggl[8\pi^2 N_c \sqrt{h_0k_0} \cr 
%&\qquad    +2\pi^2\biggl({k_0 \over h_0}h_1 + k_1\biggr)\sqrt{k_0 \over h_0}\biggl({2h_0 \over k_0}Y^2+\rho^2\biggr) 
%   +{72\pi^2 b^2 \over 5 c^2 \sqrt{h_0k_0}}{1 \over \rho^2} \biggr] 
%   +\CO(N_c^{-1}) \cr 
&= {32 \pi^2 c \over 5}N_c \zeta +{32\pi^2 c \over 25 \zeta}Y^2 \cr 
&\quad  +{16\pi^2 c \over 25 \zeta^2}\Bigl(2\zeta^5 -1 
-2a\zeta^{5/2}\sqrt{\zeta^5 -1}\Bigr) \rho^2 +{18\pi^2 b^2 \over  
\zeta}{1 \over \rho^2} +\CO(N_c^{-1}) . 
}$$ 
The critical value of $Y$ and $\rho$ minimizing the baryon mass $M$ 
is evaluated 
$$ 
Y_{\rm cr} = 0 ,\quad \rho_{\rm cr}^2 = {15 b \over 
2\sqrt{2}c}\sqrt{{\zeta \over 2\zeta^5 -1 
-2a\zeta^{5/2}\sqrt{\zeta^5 -1}}} . 
$$ 
 
Since the non-critical model has an effective 't Hooft model of 
order one, we find that in the non-critical case 
the size of the baryon is order one.

%%%%%%%%%%%%%%%%%%%%%%%%%%%%%% 
\newsec{Conclusions and discussions} 
 
We have considered the baryon sector in the non-anti-podal SS model, 
where the parameter $\zeta$ is introduced 
in addition to Kaluza-Klein mass $M_{\rm KK}$ 
and 't Hooft coupling $\lambda$. 
This model converges  to the original (anti-podal) SS model at $\zeta = 1$. 
 
The baryon mass formula \baryonmass\ has been calculated as a function of 
 $\zeta$ and $M_{\rm KK}$. 
We have compared the mass spectra with the experiment in the two ways. 
Firstly, identifying the two lowest modes with the experimental values of 
$n(940)$ and $\Delta(1232)$, we have obtained the relation \estI\ 
between $\zeta$ and $M_{\rm KK}$ 
and computed the mass spectra of $N$ and $\Delta$ baryons 
as shown in Table 2. 
The relation \estI\ implies that $M_{\rm KK}$ is bounded 
to be less than 487 MeV because of $\zeta \geq 1$ by definition. 
Secondly the baryon masses have been evaluated 
by the use of the minimal $\chi^2$ fitting. 
In this method, we can read that the upper bound of $M_{\rm KK}$ is 424.8 MeV. 
Anyway, in both cases, $M_{\rm KK}$ does not reach 949 MeV 
used in \refs{\SS, \HSS}. 
 
By following the method given by \HSS, 
we have analyzed the isoscalar, isovector and axial mean square radii, 
the isoscalar and isovector magnetic moments and the axial coupling. 
We have incorporated these physical quantities 
with the mass spectra of the baryons and $\rho$ and $a_1$ mesons, 
and compared them with the experiment. 
Then we have obtained $M_{\rm KK}$ as the functions of $\zeta$, 
which are depicted in \figmkkzeta. 
From these analyses we conclude that the $\zeta=1$ model, 
that is, the original SS model, is fitted best to the experiment. 
However, if without any justification $\zeta <1$ is permitted 
by some modification of SS model, we have found that 
the best-fitted values of $(\zeta, M_{\rm KK})$ 
are $(0.942, 997 {\rm [MeV]})$ by the use of the $\chi^2$ method. 
The physical quantities computed with these values 
are listed in Table 5 and are in good agreement with the experiment. 
Though the appropriate modification of the incorporation of baryons 
to SS model is still not clear to us, here there are two possible options: 
\item{$\bullet$} 
Since the weighted baryon vertex which is located 
at the tip of the U-shaped flavor D8-branes 
has an object with energy that scales with $N_c$, 
it might backreact on the flavor brane and also on the background geometry in such a way that the tip of the cigar  would 
be pulled down to $u_{\rm KK}^* (< u_{\rm KK})$. 
Then $\zeta$, which defined by \zcoord, can take the value 
in $\zeta \geq u_{\rm KK}^*/u_{\rm KK}$, 
where the lower bound of $\zeta$ is smaller than one. 
\item{$\bullet$} 
SS model is the dual of massless QCD. In order to put mass on the quarks, 
we need to consider the contribution of the open strings ending on 
the flavor D8-branes \refs{\BSS\DN\AK{--}\HHLY}. 
The tension of the open strings would pull up the D8-branes and 
the best-fitted value of $\zeta$, which is smaller than one, 
might be recovered to the value in $\zeta \geq 1$. 
 
We have also calculated the energy of the D4-brane wrapped in $S^4$ 
as a baryon vertex 
and analyzed its stability with respect to the location $u_B$ 
on the $u$ direction. 
In the confinement phase, the energy is monotonic on $u_B$, 
the baryon vertex is stabilized at $u_B=u_0$, that is to say, 
the baryon vertex stays at the tip of the flavor D8-branes. 
On the other hand, in the deconfinement phase, 
there appears an interesting property. 
This is caused by the balance between 
the tension of the $N_c$ open strings, 
which corresponds to the quarks of baryon, 
and the attractive force from the black hole. 
The parameter $u_T$ corresponds to temperature. 
Here we consider the behavior of the baryon vertex with respect to $u_T$ 
by fixing the tip of the D8-branes $u_0$. 
If $x_0 (=u_0/u_T)$ is larger than $x_{\rm cr}$ given by \deconfcr, 
the baryon vertex becomes stable at the tip of the D8-branes. 
If $x_0$ is smaller than $x_{\rm cr}$, 
the baryon vertex goes to the tip of the cigar background, 
which is a black hole. 
In other words, the baryon vertex can be realized 
at the tip of the D8-branes at temperatures 
lower than a critical temperature, 
but it falls down into the black hole at temperatures 
higher than the critical temperature. 
This property is similar to the chiral symmetry restoration \ASY. 
 
Finally we have commented on the single flavor model. 
It is impossible to apply the Skyrme model to the case of 
single flavor, because there does not exist a $U(1)$ instanton. 
On the other hand, in the holographic models, 
we can easily suppose the picture of the baryon vertex with single flavor. 
Though the instanton solution also plays an important role 
in our analysis of baryons, we conclude that the singular solution of 
the $U(1)$ gauge field is interpreted as the baryon.

%%%%%%%%%%%%%%%%%%%%%%%%%%%%%%  
 
\bigbreak\bigskip\bigskip\centerline{{\bf Acknowledgments}}\nobreak 
 
We would like to thank Ofer Aharony and Vadim Kaplunovsky 
for useful conversations and their comments on the manuscript.
JS would like to thank Elias Kiritsis and Shigeki Sugimoto for 
frutiful discussion during the Institute d'ete, Ecole Normal. 
This work was supported in part by a center of excellence
supported by the Israel Science Foundation (grant number 1468/06), 
by a grant (DIP H52) of the German Israel Project Cooperation, 
by a BSF United States-Israel binational science foundation grant 2006157   
and by the European Network MRTN-CT-2004-512194. 
The work of JS was also supported in part by European Union
Excellence Grant MEXT-CT-2003-509661.
SS are grateful to Institute of Physical and Chemical Research
(RIKEN) and Yukawa Institute for Theoretical Physics (YITP) 
for their hospitality. 
Part of the work was done while SS was visiting YITP 
with partial support by the Grant-in-Aid for the Global COE Program 
``The Next Generation of Physics, Spun from Universality and
Emergence" 
from the Ministry of Education, Culture, Sports, Science and
Technology (MEXT) of Japan.

%%%%%%%%%%%%%%%%%%%%%%%%%%%%%%  
\appendix{A}{baryon vertex in D$p$-branes' background} 
 
Let us consider the energy $E_p$ of D$(8-p)$-brane wrapped on $S^{8-p}$ and $N_c$ fundamental strings, which is denoted by 
$$ 
S_p = -T_{8-p}\int dt d\Omega_{8-p} e^{-\phi}\sqrt{-\det g_{{\rm D}(8-p)}} 
    -N_cT_f\int dt du \sqrt{-\det g_{\rm string}} 
    =: \int dt E_p , 
$$ 
where the tension of D$(8-p)$-brane $T_{8-p} = (2\pi)^{p-8}l_s^{p-9}$.

%%%%%%%%%%%%%%%%%%%%%%%%%%%%%% 
\subsec{Confinement phase} 
 
The metric of the background is described as 
$$\eqalign{ 
&ds^2 = \biggl({u \over R_p}\biggr)^{{7-p \over 2}}\biggl[-dt^2 +\sum_{i=1}^{p-1} (dx^i)^2 + f(u;p)(dx^p)^2 \biggr] 
    +\biggl({R_p \over u}\biggr)^{7-p \over 2}\biggl[{du^2 \over f(u;p)} +u^2d\Omega_{8-p} \biggr] , \cr 
&R_p^{7-p} = {g_sN_c(2\pi l_s)^{7-p} \over (7-p)V_{8-p}} ,\quad 
e^\phi = g_s\biggl({R_p \over u}\biggr)^{(7-p)(3-p) \over 4} ,\quad 
f(u;p) = 1-\biggl({u_\Lambda \over u}\biggr)^{7-p} , 
}$$ 
where $V_{8-p}$ is the unit volume of $S^{8-p}$, which is equal to $2\pi^{(9-p)/2}/\Gamma((9-p)/2)$. 
The energy is described as 
$$\eqalign{ 
E_p(u_B;u_0) &= {N_c u_\Lambda \over 2\pi l_s^2} {\cal E}^{(p)}_{\rm conf} ,\quad 
{\cal E}^{(p)}_{\rm conf}(x;x_0) = {1 \over 7-p}x +\int_x^{x_0}{dy \over \sqrt{1-y^{p-7}}} ,\cr 
&x:= {u_B \over u_\Lambda},\quad x_0 := {u_0 \over u_\Lambda} ,\quad 1 \leq x \leq x_0. 
}$$ 
The integration can be computed in terms of the hypergeometric function ${}_2F_1$, 
$$\eqalign{ 
\int_x^{x_0}{dy \over \sqrt{1-y^{p-7}}} = 
&-{2ix^{9-p \over 2} \over 9-p} {}_2F_1\biggl({p-9 \over 2p-14},{1 \over 2},{23-3p \over 14-2p},x^{7-p}\biggr) \cr 
&+{2ix_0^{9-p \over 2} \over 9-p} {}_2F_1\biggl({p-9 \over 2p-14},{1 \over 2},{23-3p \over 14-2p},x_0^{7-p}\biggr) . 
}$$

%%%%%%%%%%%%%%%%%%%%%%%%%%%%%% 
\subsec{Deconfinement phase} 
 
The metric of the background is described as 
$$\eqalign{ 
&ds^2 = \biggl({u \over R_p}\biggr)^{{7-p \over 2}}\biggl[-f_T(u;p) dt^2 +\sum_{i=1}^p (dx^i)^2 \biggr] 
    +\biggl({R_p \over u}\biggr)^{7-p \over 2}\biggl[{du^2 \over f_T(u;p)} +u^2d\Omega_{8-p} \biggr] , \cr 
&R_p^{7-p} = {g_sN_c(2\pi l_s)^{7-p} \over (7-p)V_{8-p}} ,\quad 
e^\phi = g_s\biggl({R_p \over u}\biggr)^{(7-p)(3-p) \over 4} ,\quad 
f_T(u;p) = 1-\biggl({u_T \over u}\biggr)^{7-p} . 
}$$ 
The energy $E_p$ can be evaluated, 
$$\eqalign{ 
E_p(u_B;u_0) &= {N_c u_T \over 2\pi l_s^2}{\cal E}^{(p)}_{\rm deconf} ,\quad 
{\cal E}^{(p)}_{\rm deconf}(x;x_0) = {1 \over 7-p}x\sqrt{1-x^{p-7}} + (x_0 - x) ,\cr 
&x:= {u_B \over u_T},\quad x_0 := {u_0 \over u_T} ,\quad 1 \leq x \leq x_0. 
}$$ 
\bigskip 
\vbox{\centerline{\epsfbox{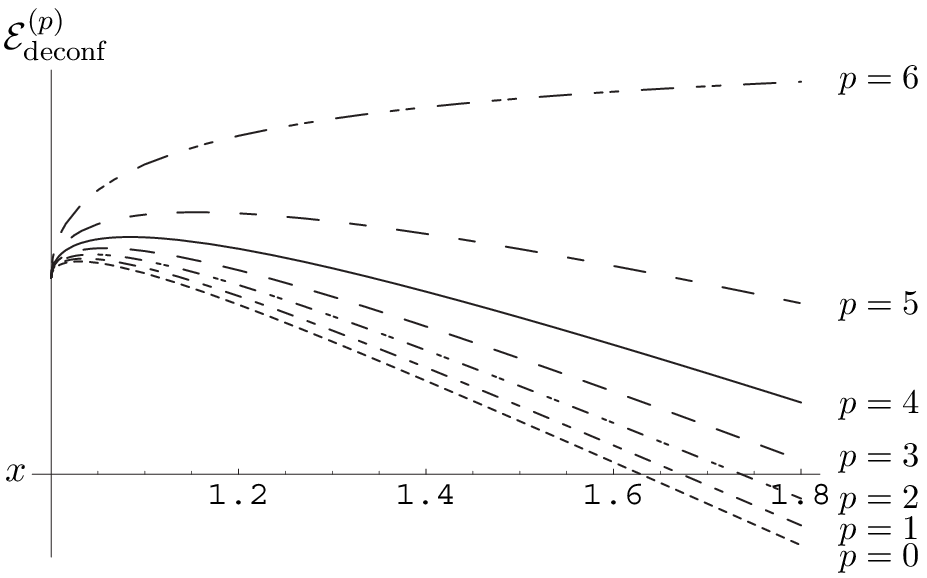}} 
\centerline{\fig\figpdeconf{${\cal E}^{(p)}_{\rm deconf}(x)$} ${\cal E}^{(p)}_{\rm deconf}(x)$}} 
\bigskip\nobreak\noindent 
Fig.~\xfig\figpdeconf\ implies that only $E_6(u_B)$ is a monotonically increasing function.

\listrefs 
 
\bye